\documentclass[twocolumn,showpacs,floatfix,pre]{revtex4-1}
\RequirePackage{lineno}

\usepackage{graphicx}
\usepackage{graphicx}
\usepackage{dcolumn}
\usepackage{epsf}
\usepackage{amsmath}
\usepackage{bm}
\usepackage{setspace}
\newcommand{\bcc}{{\it bcc\,\,}}

\newcommand{\pdf}{{\it pdf\,}}
\newcommand{\wave}{{\it wave\,}}

\newcommand{\sscf}{{\it sscf\,\,}}
\newcommand{\sscfs}{{\it sscf}s\,\,}
\newcommand{\pbc}{{\it PBC\,\,}}
\newcommand{\auto}{{\it auto}}
\newcommand{\self}{{\it self\,\,}}
\newcommand{\cross}{{\it cross\,\,}}
\newcommand{\total}{{\it total\,\,}}

\newcommand{\tccf}{{\it tccf\,\,}}

\newcommand{\tongue}{{\it tongue\,\,}}
\newcommand{\bonfire}{{\it bonfire\,\,}}
\newcommand{\sickle}{{\it sickle\,\,}}

\begin{document}


\title{Dependence of the Atomic Level Green-Kubo Stress Correlation Function on 
Wavevector and Frequency. Molecular Dynamics Results from a Model Liquid.} 

\author{V.A. Levashov}
\affiliation{Department of Physics and Astronomy, University of
Tennessee, Knoxville, TN 37996, USA.}


\begin{abstract}
{
We report on a further investigation of a new method that can be used to address vibrational 
dynamics and propagation of stress waves in liquids. 
The method is based on the decomposition of the macroscopic Green-Kubo stress correlation 
function into the atomic level stress correlation functions. 
This decomposition, as was demonstrated previously for a model liquid studied in 
molecular dynamics simulations, reveals the presence of stress waves propagating over 
large distances and a structure that resembles the pair density function. 
In this paper, by performing the Fourier transforms of the atomic level stress correlation functions, 
we elucidate how the lifetimes of the stress waves and the ranges of their 
propagation depend on their frequency, wavevector, and temperature. 
These results relate frequency and wavevector dependence of the generalized viscosity to 
the character of propagation of the shear stress waves.  
In particular, the results suggest that an increase in the value of
the frequency dependent viscosity at low frequencies with decrease of temperature
is related to the increase in the ranges of propagation of the stress waves of the corresponding low frequencies.  
We found that the ranges of propagation of the shear stress waves of frequencies less 
than half of the Einstein frequency, extend well beyond the nearest neighbor 
shell even above the melting temperature. 
The results also show that the crossover from quasilocalized to 
propagating behavior occurs at frequencies 
usually associated with the Boson peak.
}
\end{abstract}

\today

\maketitle


\section{Introduction}\label{s:intro}

In molecular dynamics (MD) simulations, the dependence of the generalized viscosity on frequency and wavevector 
is often studied using the transverse current correlation function (\tccf)
\cite{HansenJP20061,Boon19911,EvansDJ19901,Levesque19731,EvansDJ19811,Alder19831,
Alley19831,Mountain19821,PalmerBJ19941,Kaufman2007E,
Puscasu110,Tanaka20091,Tanaka2011E,Furukawa2013E,Mizuno2012,Mizuno2013}.
It has been demonstrated for the \tccf  and the generalized viscosity that 
upon decrease of temperature toward the glass transition there occurs a significant 
increase in their values for small wavevectors, 
i.e., for the wavelengths larger than the lengths associated with the second coordination shell
\cite{Alder19831,PalmerBJ19941,Puscasu110,Keyes20051,Tanaka20091,Tanaka2011E,Furukawa2013E}.
However, there is not a large increase for the wavevectors associated with 
the first coordination shell and larger wavevectors 
\cite{Alder19831,PalmerBJ19941,Puscasu110,Tanaka20091,Tanaka2011E,Furukawa2013E}.
Thus, results for the \tccf differ from the results for the structural relaxation, which
is often studied with the intermediate scattering function. Relaxation time
for the intermediate scattering function, is usually 
determined (due to de Gennes narrowing) using
the value of wavevector corresponding to the nearest neighbor distance.
It has also been shown that properties of the \tccf
function can be modeled using kinetic and viscoelastic models 
if  it is assumed that transport coefficients depend on the value of wavevector \cite{Alley19831,Mizuno2012,Mizuno2013}.
These results suggest a non-local nature of the \tccf and viscosity 
close to the glass transition temperature.

A different, but a closely related approach for understanding viscosity
is based on the Green-Kubo expression and considerations of 
the macroscopic stress-stress correlation function (\sscf) 
\cite{HansenJP20061,Boon19911,EvansDJ19901,Levesque19731,EvansDJ19811,
Alder19831,Kaufman2007E,Tanaka2011E,Tang1995,Green1954,Kubo1957,Helfand1960}.
The Green-Kubo expression for viscosity corresponds to zero-wavevector ($\bm{k}=0$) 
and zero-frequency ($\omega=0$) limit of the expression for generalized viscosity 
\cite{Green1954,Kubo1957,Helfand1960,HansenJP20061,Boon19911,EvansDJ19901,EvansDJ19811,Alder19831}.

The Green-Kubo method is very common in MD simulations \cite{Hoheisel19881,Meier20041}. 
However, {\it the microscopic} nature of the {\it macroscopic} stress correlation function 
is poorly understood. 
Sometimes it was assumed that the atomic level \sscf is local
and that there are correlations between the nearest neighbor 
atoms only \cite{Chialvo19931,Allen19941,MCQuarrie19761}.
This view contradicts the older and more recent results from
generalized hydrodynamics \cite{HansenJP20061,Boon19911,EvansDJ19901,
Levesque19731,EvansDJ19811,Alder19831,
Mountain19821,PalmerBJ19941,Kaufman2007E,
Puscasu110,Tanaka20091,Tanaka2011E,Furukawa2013E,
Mizuno2012,Mizuno2013,Todd20081,Todd20082,
Schepper1987,Schepper1988,Mryglod1995,Mryglod19971,
Mryglod19972,Mryglod19973,Mryglod20051,Bertolini20111}.

Previously we studied the {\it microscopic} nature 
of the {\it macroscopic} Green-Kubo 
\sscf by decomposing it into correlation functions between 
the local atomic level stresses \cite{Levashov2013,Levashov20111}.
Our results explicitly demonstrate non-locality of the stress correlations. 
They also show that there is a relation between propagating 
transverse (shear) waves and viscosity.
In this paper, we assume that the reader is familiar with the results 
presented in Ref.\cite{Levashov2013,Levashov20111}.

Recently, the non-local correlated character of the particles' displacements in model liquids was  
discussed in the context of the Eshelby field \cite{Shall20111,Picard20041,Lemaitr20131}.
There have also been observations made concerning propagating longitudinal and transverse waves \cite{Lemaitr20131}. 
Non-locality of the correlations is reflected in recent theories \cite{Mirigian20131}. 

In order to get intuitive insight  into the connection between 
the propagation of waves and the atomic level \sscf we considered 
a simple model in Ref.\cite{Levashov20141}. 
In this model propagating waves are plane waves, like in crystals. 
It is also assumed there that the atomic environment of every atom is spherically symmetric. 
We found that if an additional assumption concerning the decay of the stress correlation 
function for a given frequency is introduced, the atomic level \sscf calculated within
this toy model qualitatively resembles the atomic level \sscfs that were obtained in MD 
simulations on a model liquid in Ref.\cite{Levashov2013,Levashov20111}. 
While it is clear that this model cannot be used to describe liquids, as in liquids vibrational eigenmodes are not plane waves,
it still provides insight into the atomic level \sscf in liquids. 
The results that were obtained in Ref.\cite{Levashov20141} with respect to the Fourier transforms 
provide a guide for the present analysis.

In this paper, by performing the Fourier transform of the previous MD data \cite{Levashov2013,Levashov20111}, 
we demonstrate how the atomic level \sscf method can be used
to study properties of the stress waves and how these properties depend on frequency and wavevector. 
The results show that high-frequency stress waves are quasilocalized
and temperature decrease does not strongly affect the degree of their localization.
At the same time high-frequency viscosity exhibits only weak temperature dependence. 
However, the ranges of propagation of low-frequency stress waves significantly increase with decrease of temperature.
This increase correlates with the significant increase in the value of low-frequency viscosity.
Our data also show that the change from quasilocalized to propagating behavior happens in the range of frequencies  
associated with the Boson peak, as expected \cite{Taraskin20021,Tanaka20081,Ruocco20131,Ediger20121}.

In this paper, we introduce a wavevector $\bm{q}$ that characterizes
the length scales relevant to the atomic level Green-Kubo \sscf and to the propagating shear stress waves. 
It is important to realize that the wavevector $\bm{q}$
is distinct from the wavevector $\bm{k}$ in the \tccf approach. 
All our results correspond to the case of $\bm{k}=0$,  
i.e., to the case of density fluctuations of very large wavelengths.
This effectively means that we study shear stress waves in the absence of local 
density fluctuations.
See Appendix \ref{ax:tccf} for more details.

The paper is organized as follows.
In section \ref{s:mddetails} we provide some details
about our MD system. In section \ref{s:separation} we describe
how certain features in the \sscf can be separated. 
This separation is useful for further analysis.
In section  \ref{s:wwsee} we discuss some 
features of the \sscf.
In section  \ref{s:ftrans} we describe how we apply the 
Fourier transforms.
In section \ref{s:results1500} we describe the results of the Fourier
transforms at the lowest temperature that we studied.
In section \ref{s:etaw} we discuss the connection between our results
and frequency dependent viscosity.  
In section \ref{s:Tevol} we discuss the results of the 
Fourier transforms at several higher temperatures.
We conclude in section \ref{s:conclusion}.
In Appendix \ref{ax:tccf} we discuss the relation 
of our approach to the \tccf approach.
In Appendices \ref{s:bonfire} and \ref{s:tongue}
we address two particular features of the Fourier transforms. 

\section{Details of MD simulations and Reduced Units \label{s:mddetails}}

MD simulations has been performed in NVT ensemble on a single component system of
particles that is supposed to mimic liquid iron. 
The number density of the system corresponded
to a \bcc lattice with lattice spacing $a =2.9434$ (\AA). The
particles interact through a short range pairwise potential.
The potential crosses zero at
$r_{\sigma} =2.2245$ (\AA). It has a minimum at 
$r_{min}\approx 2.6166$ (\AA) with
the depth $\epsilon \approx 0.2516$ (eV). 
This depth approximately corresponds to the temperature 2800 K. 
The potential is zero beyond 3.44 (\AA). 

The potential energy landscape crossover temperature of this 
system is $T_A \approx 2300$ (K) \cite{Levashov2008E}. 
The mode coupling temperature is $T_{MC} \approx 1150$ (K), 
and the glass transition temperature is $T_g \approx 950$ (K).

A fifth order Gear predictor-corrector algorithm was
used to integrate equations of motion with time step  
$10^{-15}$ (s), i.e., 1 Femtosecond (fs) 
for $T \leq 5000$ (K) and 0.5 (fs) for
$T > 5000$ (K). 
Temperature was introduced through rescaling of the particles velocities.

Our MD simulations have been described in more details in Ref. \cite{Levashov2013,Levashov20111}.
Previously, when reporting results of MD simulations, 
we had not used reduced units.
To be consistent, we also used non-reduced units in this paper. 
To make a comparison between our results
and the results from other publications, 
we describe in Ref.\cite{reduced} the relations between our units
and the units of the corresponding Lennard-Jones potential.
 
\section{Separation of the \sscf into the \pdf-like and \wave-like  parts  \label{s:separation}}

\begin{figure*}
\begin{center}
\includegraphics[angle=0,width=6.0in]{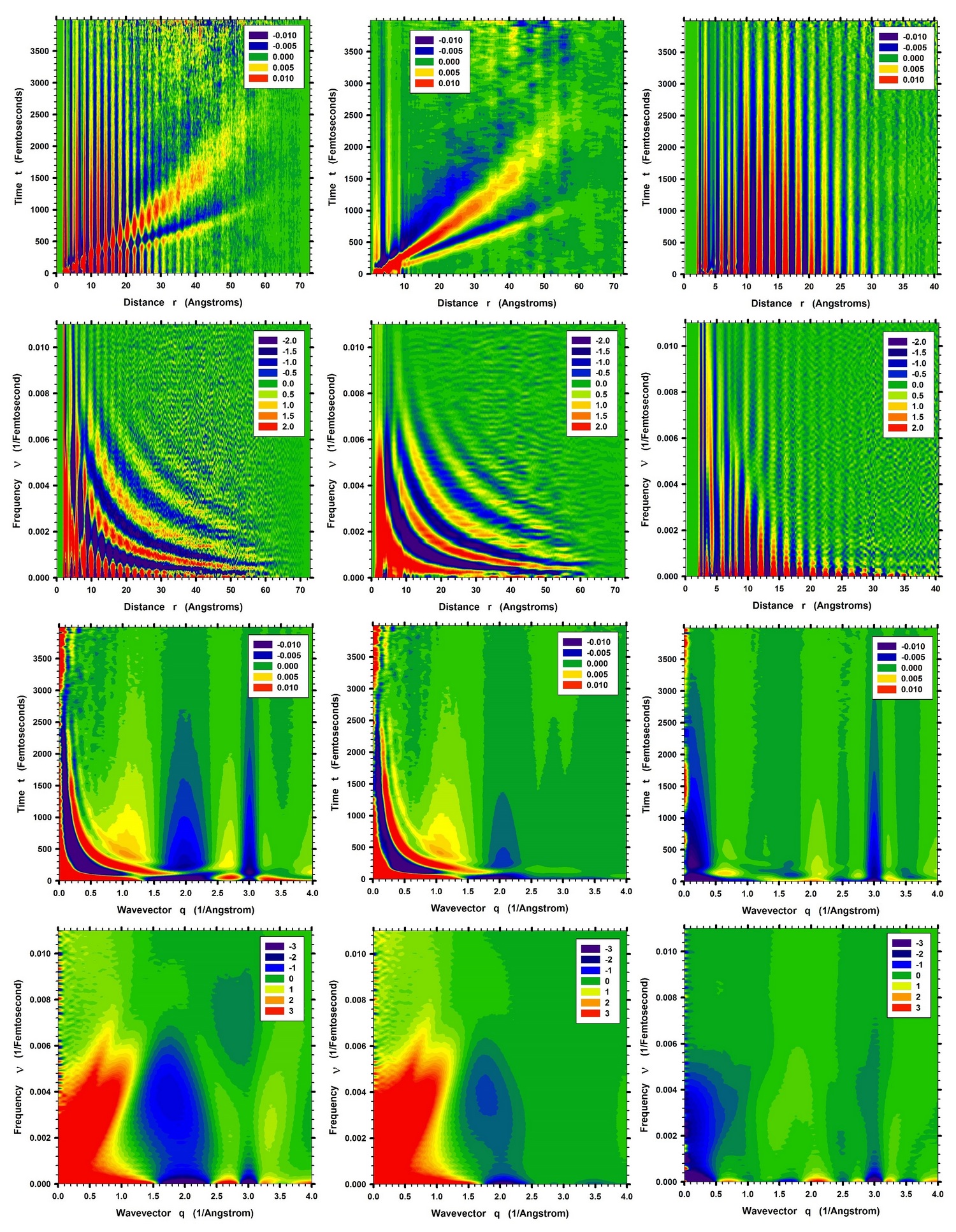}
\caption{The \sscf of the large (43904) system at 1500 K, the \wave  and \pdf-like contributions
to it, and their Fourier transforms. The notation $(n,m)$ in the text will be used to
refer to the panels of the figure: $n$-for the rows and $m$-for the columns.
One should not consider the values of $q<0.153 \AA^{-1}$ and the values 
of $\omega<0.00025 (fs)^{-1}$.
}\label{fig:Fig1}
\end{center}
\end{figure*}

\begin{figure}
\begin{center}
\includegraphics[angle=0,width=3.4in]{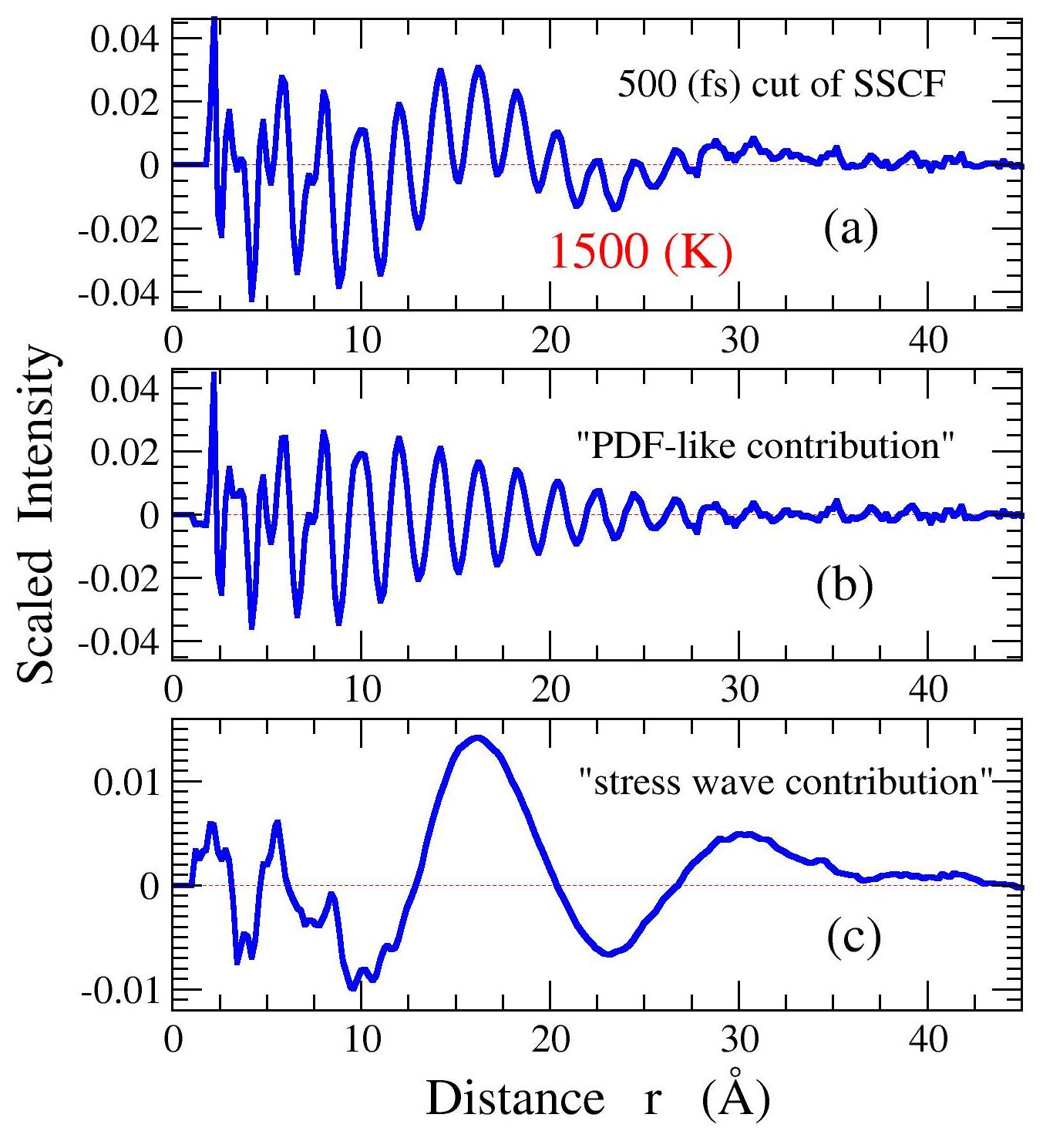}
\caption{ 
Separation of the 500 (fs) cut of the SSCF at 1500 (K) into the \pdf-like 
and \wave-like contributions. 
(a) 500 (fs) cut of the SSCF from panel (1,1) of Fig.\ref{fig:Fig1}. 
(b) \pdf-like contribution. (c) Contribution from stress waves.  
}\label{fig:Fig02}
\end{center}
\end{figure}

Some figures in this paper have several panels.
In referring to these panels we use $(n-row,m-column)$ notation. 

Panel (1,1) of Fig.\ref{fig:Fig1} shows the \sscf 
at 1500 (K) obtained on the large system of 43904 particles 
with $(L/2) \approx 41.21 $ (\AA) \cite{Levashov2013}, where
$L$ is the length of the side of the cubic box.
The main features present in the \sscf are the stress waves and 
the pair distribution function \pdf-like stripe structure.
To understand the results of the Fourier transforms of this panel, it is convenient 
to make an approximate separation of the \sscf into the parts that correspond 
to the waves and to the \pdf-like structure. 

The implemented separation procedure is  
based on the following observation. 
Consider the zero-time cut of panel (1,1) of Fig. \ref{fig:Fig1}.
This cut is shown as the lowest curve in Fig.6 of Ref.\cite{Levashov2013}.
Note the oscillating behavior for the distances beyond 10 (A).
Also note that for a particular maximum, which is beyond 10 \AA, 
the distances from it to the nearest two minimums are approximately the same. 
Thus, for a particular maximum at $r$ the value of the \sscf 
at it, $f_{gr}(r)$, could be approximated as:
\begin{eqnarray}
f_{gr}(r) \approx - \frac{1}{2}\left[f_{gr}(r-\Delta_r) + f_{gr}(r+\Delta_r)\right]\;\;,
\label{eq;separate-01}
\end{eqnarray}
where  $f_{gr}(r-\Delta_r)$ and $f_{gr}(r+\Delta_r)$ 
are the values of the \sscf at the left and right minimums nearest to this maximum. 
Finally, note that (\ref{eq;separate-01}) can be used not only for the maximum/minimum values, 
but essentially for all $r$ beyond the third or forth coordination shells. 
This, however, does not hold for nonzero times and those regions of $r$ that contain 
contributions from the waves, as can be seen in Fig.\ref{fig:Fig02}(a). 

Further, we assume that the \pdf-like contribution to the 
\sscf satisfies (\ref{eq;separate-01}), while 
the wave's contribution does not. 
In our iterative numerical procedure we consider different times independently. 
We assume that for a particular time we know the distance dependence 
of the wave contribution on step $n$, i.e., $\bar{f}^n_w(r)$. 
On the first step we assume that it is zero for all distances.
Then we calculate the \pdf-like contribution to the \sscf:
\begin{eqnarray}
f_{gr}^n(r) = f(r) -\bar{f}^n_w(r)\;\;.
\label{eq;separate-02}
\end{eqnarray}
Then, in accord with (\ref{eq;separate-01}), 
we define:
\begin{eqnarray}
f_{gr}^{n+1}(r) = - \frac{1}{2}\left[f_{gr}^n(r-\Delta_r) + f_{gr}^n(r+\Delta_r)\right]\;\;,
\label{eq;separate-03}
\end{eqnarray}
where $\Delta_r$ is the distance from the nearest maximum to its nearest minimum. 
$\Delta_r$ is distance dependent (this dependence is weak in practice). 
Then we define the contribution from the wave on step $n+1$ as:
\begin{eqnarray}
f_{w}^{n+1}(r) = f(r)-f_{gr}^{n+1}(r)\;\;.
\end{eqnarray}
Finally, we assume that the amplitude of the wave does not change significantly 
over the distance between the nearest maximum and minimum. 
It is indeed so according to Fig.\ref{fig:Fig02}.
Thus, for convergence of the algorithm, we introduce the average amplitude 
of the wave, $\bar{f}_w^{n+1}$, in which the averaging 
goes over the interval $(r-\Delta_w,r+\Delta_w)$:
\begin{eqnarray}
\bar{f}_{w}^{n+1}(r) =\frac{1}{2\Delta_w}\int_{r-\Delta_w}^{r+\Delta_w} f_w^{n+1}(\xi)d\xi\;\;
\label{eq;separate-05}
\end{eqnarray}
With this new value of $\bar{f}_{w}^{n+1}(r) $ we go back to (\ref{eq;separate-02}) 
closing the iteration loop.
\begin{figure}
\begin{center}
\includegraphics[angle=0,width=3.4in]{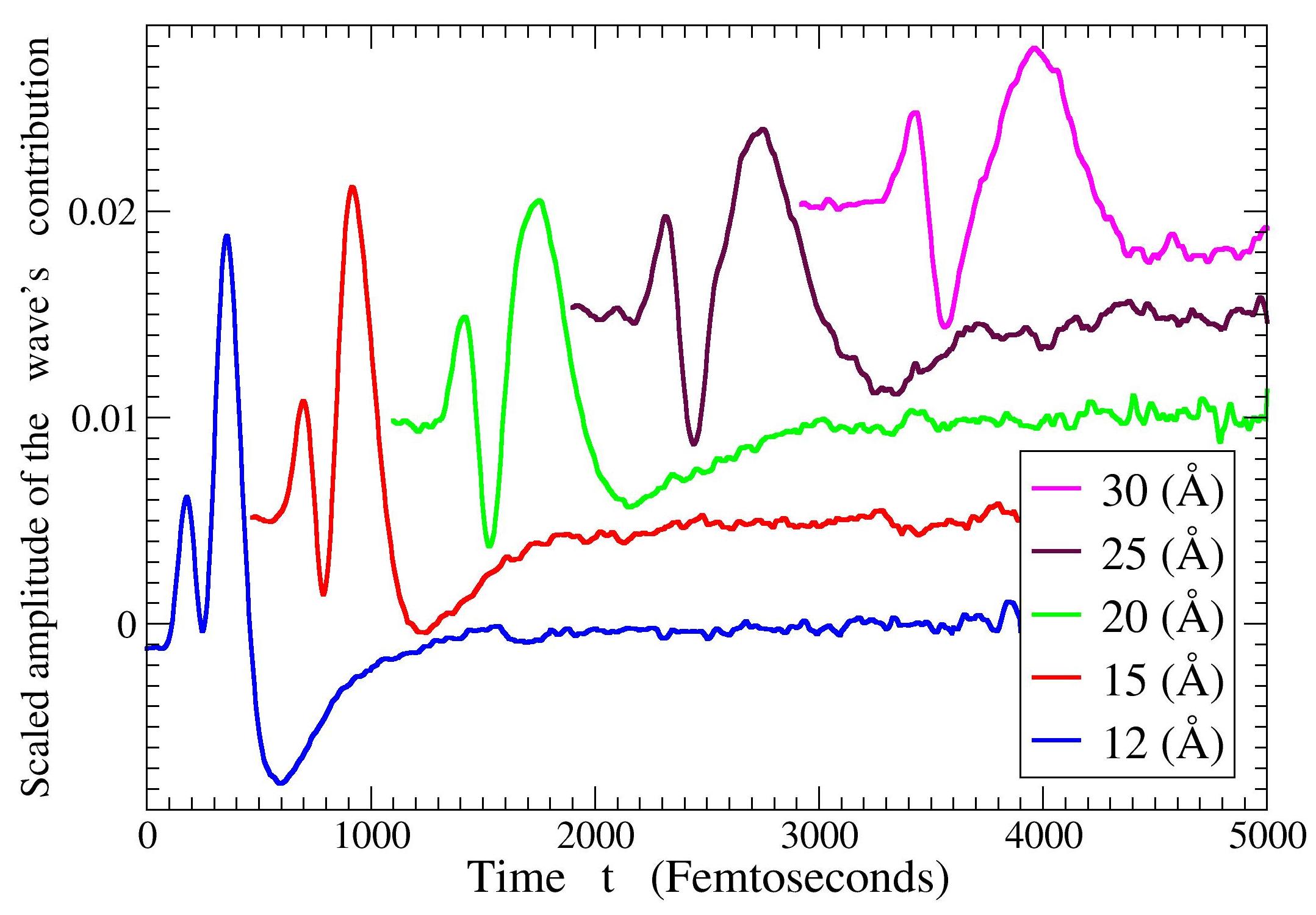}
\caption{Constant $r$ cuts of panel (1,2). 
The legends show the values of $r$. 
The curves were shifted so that they do not overlap.
The first peaks in the curves are due to the compression wave, 
while the second peaks are due to the shear wave.
Note that the shear peaks are larger and broader.
The scale on the $y$-axis, corresponds to the correlation 
between the central atom and atoms in the spherical annulus 
of radius $r$ and thickness $dr = 0.2$  (\AA). 
Normalization to the atomic stress \auto-correlation functions was also made.
}\label{fig:Fig03}
\end{center}
\end{figure}

The results of the described procedure are shown in Fig.\ref{fig:Fig02}
for temperature 1500 (K) and for the time-cut 500 (fs).
The values of $\Delta_r$ were extracted from zero-time cuts of the \sscf, 
while the value used for $\Delta_w =1$ (\AA).
The number of iterations was 1000.

Note that the procedure does not work
for distances $r<10$ (\AA). This is so because for such $r$ in the \pdf-like contribution 
there is no periodicity assumed in the separation procedure. 
The value of $r=10$ (\AA) approximately corresponds to 
the inclusion of the 4th coordination shell.
This distance could be associated with the medium range order distance \cite{Sokolov19921,Tomida19951}.
It is of interest that viscoelastic continuous approximations appear to be valid at distances 
larger than $10$ (\AA), but not at smaller distances \cite{Mizuno2013}.

We applied the described algorithm to all times in panel (1,1) of Fig.\ref{fig:Fig1}. 
The results are shown in panels (1,2) and (1,3). 
Note again that the procedure does not work for $r<10$ (\AA).
Figure \ref{fig:Fig03} shows constant $r$ cuts from
panel (1,2). 
First peaks in the curves correspond to the compression wave,  
the second peaks to the shear wave.
It would be useful to develop a method that would allow 
us separate contributions from the shear and compression waves.

\section{What we see in panels \\ (1,1), (1,2), and (1,3) \label{s:wwsee}}

The intensity in panel (1,1) of Fig.\ref{fig:Fig1}
shows the ensemble averaged atomic level stress correlation 
function between a central atom and atoms
located inside the spherical annulus of radius $r$ 
and thickness $\Delta r = 0.2$ (\AA). 
This intensity is also normalized to the magnitude of the 
stress \auto-correlation function at zero time 
\cite{Levashov20111,Levashov2013}. 
Figures 6 and 7 of Ref.\cite{Levashov2013}, 
and Figures \ref{fig:Fig02},\ref{fig:Fig03} 
of this paper further clarify the scale of the correlations.
For example, the magnitude of the stress correlation of a central
atom with the maximum intensity annulus in the first coordination shell  
at zero time is $\sim 0.35$ of the stress \auto-correlation function at zero time 
(width of the annulus is $\Delta r=0.2$ (\AA)). 
For the maximum intensity annulus in the second coordination shell this ratio is less than $0.10$
(from Fig.6 of Ref.\cite{Levashov2013}).
It is also useful to recall that the value of the stress \auto-correlation 
function at zero time determines the value of the atomic 
level stress energy \cite{Egami19821,Chen19881,Levashov2008B}.

\begin{figure}
\begin{center}
\includegraphics[angle=0,width=3.0in]{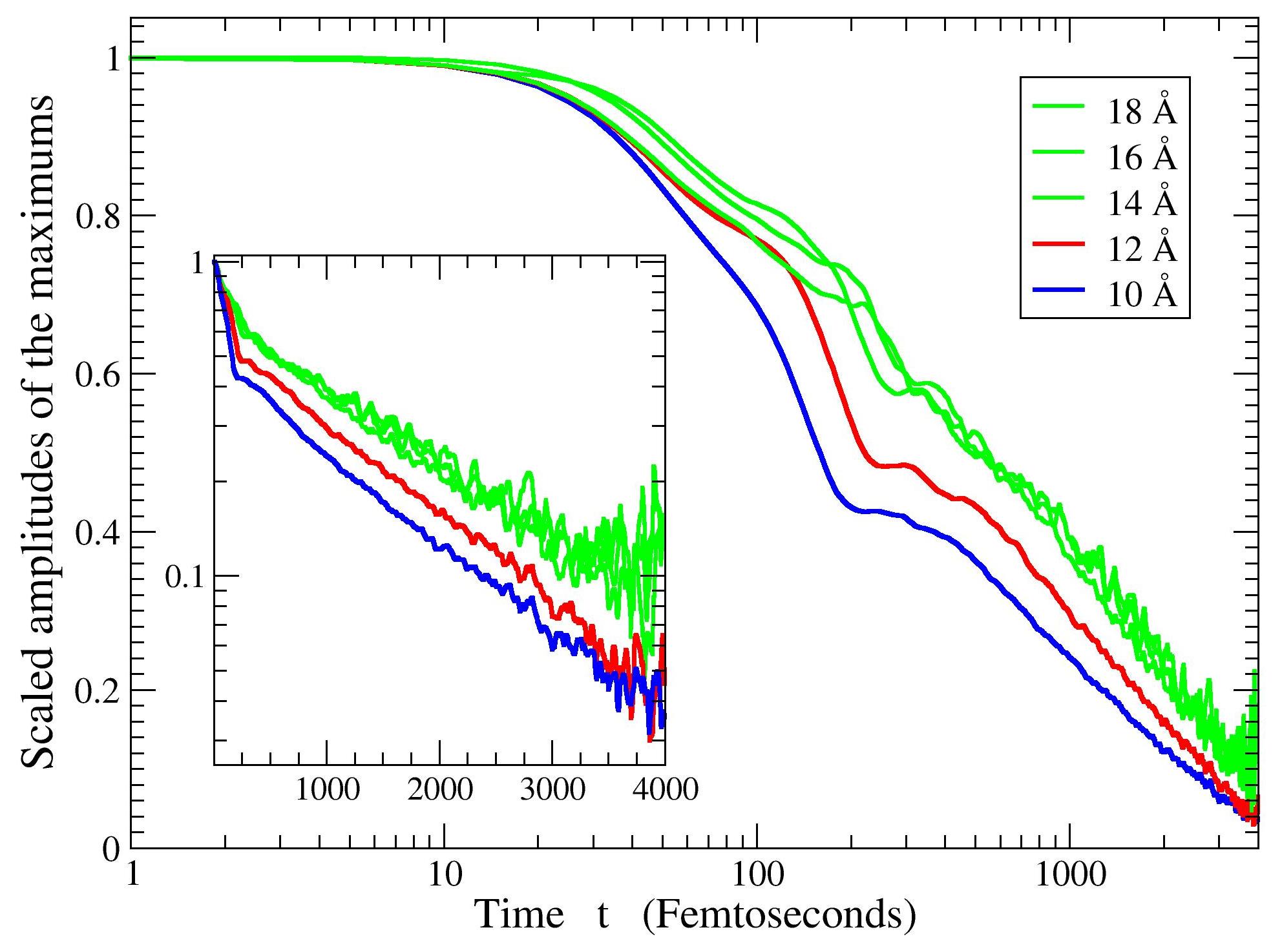}
\caption{ 
Dependencies on time of the stripes' maximums 
in the \pdf-like contribution to the \sscf. 
See panel (1.3) of Fig.\ref{fig:Fig1}.
The legends show the positions of the stripes.
The figure and the inset show the same data.
Linear behavior of the curves in the inset at large times
suggests a relation between the decay of the maximums 
of the stripes and diffusion.
}\label{fig:Fig04}
\end{center}
\end{figure}
Panels (1,1), (1,2), and (1,3) raise several
questions. For example, as panel (1,1) shows the atomic level decomposition
of the macroscopic {\it shear} \sscf
$\left<S^{xy}S^{xy}\right>$, 
it is reasonable to wonder why we see in 
it the \pdf-like structure and the contributions 
from the longitudinal waves.
Indeed, both of these features should be 
related to the density fluctuations and 
not to the {\it shear} \sscf. 
The explanation can be related to the results presented 
in Ref.\cite{Kust2003a,Kust2003b}.
There it was shown that different components of the atomic level
stress tensor on the same atom are correlated.
On the other hand, the presence of correlations between 
the different stress components on the same atom
questions the results and 
derivations in Ref.\cite{Egami19821,Chen19881,Levashov2008B}, 
as there, when the {\it equipartition} law is derived, it is assumed 
that different stress components on the same atom are independent. 
However, the derivations of the equipartition in  \cite{Egami19821,Chen19881,Levashov2008B} 
are based on the Taylor expansion and considerations of only those terms which 
are quadratic in atomic strain.
Under this quadratic approximation, different stress components on the same atom are independent 
in the spherical (cubic) representation.
Thus, the results presented in Ref.\cite{Kust2003a,Kust2003b} can be related to the higher 
order terms in the Taylor expansion.
This means that the density-density correlations which we see in 
panels (1,1), (1,2), and (1,3) can be related to higher order terms.
These questions require further clarifications. 

\subsection{\pdf-like contribution}

It is reasonable to expect that the decay of the 
\pdf-like contribution in panel (1,3) is related to the decay
of the van Hove correlation function \cite{vanHove}.
Figure \ref{fig:Fig04} shows dependencies on time of the maximums in the \pdf-like
contribution to the \sscf. For example, 12 (\AA) maximum was found as the maximum value
(for every time) of the \pdf-like contribution in the 
interval of distances between 11 (\AA) and 13 (\AA). 
Differences between the 10 (\AA), 12 (\AA) and the other curves are likely to be caused
by the stress waves: as intensities of the stress waves 
are larger at small distances it is likely that these higher intensities 
stimulate faster decay in the \pdf-like stripes.
The decay in the amplitudes of the stripes for $t < 200$ (fs)
is likely to be due to the rattling cage motion. 
The decay for $t > 300$ (fs) is likely to be related to the particle diffusion.
If we assume that the particles that diffuse away from the spherical annulus
completely lose the correlation with the original state, 
while those that remain keep this correlation, then the magnitude of 
the remaining correlation should be proportional
to the number of the particles remaining in the annulus. 
Since the rate of diffusion away from the spherical annulus should be proportional
to the number of particles remaining in the annulus, 
the number of the remaining particles should decrease
exponentially with time. Thus the \pdf-like stress correlation function at large distances
at $t>300$ (fs) should decay exponentially with time. 
This behavior can be observed in the inset of Fig.\ref{fig:Fig04}.

\subsection{Stress waves' contribution}

It is clear that panel (1,2) shows propagating
shear and compression waves. 
Previously we argued that shear stress waves 
are related to viscosity \cite{Levashov20111,Levashov2013}.
Thus, it is important to understand the features in panel (1,2).
However, it is not clear {\it how} stress waves translate into the
features observed in the \sscf. The nature of  these stress waves also remains obscure. 
These are complicated questions for liquids, as currently there is no accepted and convenient
way to describe vibrational dynamics in disordered media and its coupling 
to diffusion \cite{Keyes19971,Taraskin20001,Taraskin20021,Ediger20121,Tan20121}.
In our view, it is possible that the atomic level stress correlation 
function that we consider here represents an alternative way to describe vibrational dynamics.

In order to gain at least some insight into the connection between the vibrational 
dynamics and the atomic level stress correlation function 
we considered in Ref.\cite{Levashov20141}
a simple model. In this model vibrational modes are represented by plane waves, like in crystals.
Of course, plane waves do not represent vibrational eigenmodes 
of liquids \cite{Taraskin20021,Ediger20121,Keyes19971,Taraskin20001,Tan20121}.
However, in our view, considerations in Ref.\cite{Levashov20141} provide insight 
into the nature of the connection between the stress waves and the atomic level stress correlation function.

\section{Fourier Transforms of the \sscf  \label{s:ftrans}}

In our previous considerations, the atomic level \sscf, $F(t,r)$, 
is defined as a correlation function
between a central atom and atoms inside the spherical annulus of radius $r$ and thickness
$dr$ \cite{Levashov20111,Levashov2013}. 
This definition naturally follows from the Green-Kubo expression for viscosity.
For further analysis and in view of Ref.\cite{Levashov20141} 
we introduce:
\begin{eqnarray}
f_p(t,r) \equiv \frac{1}{(4\pi r^2)} F(t,r)\;,\;\;\;\;\;f_r(t,r) \equiv r \cdot f_p(t,r)\;\;,\;\;\;
\label{eq;fpandfr}
\end{eqnarray}
where  $f_p(t,r)$ is the atomic level stress correlation
per pair of particles.

We define the Fourier transform over $t$ of $f_r(t,r)$ as:
\begin{eqnarray}
\tilde{f}_r(\omega,r)\equiv \int_0^{\infty}f_r(t,r)\cos(\omega t) dt\;\;.
\label{eq;ttow}
\end{eqnarray}
It was shown, in the framework of the model discussed in Ref.\cite{Levashov20141}, 
that if vibrations are non-decaying plane waves,
then $\tilde{f}_r(\omega,r)$ should, for every $\omega$, exhibit {\it constant amplitude}
oscillations in $r$ with a wavelength determined by the dispersion relation $\omega(q)$.
Since, for different $r$ the Fourier transforms over $t$ are independent, we
transform $F(t,r)$ instead of $f_r(t,r)$. 
In the case of non-decaying plane waves, amplitudes of peaks in
$\tilde{F}(\omega,r)$ should linearly increase with increase of $r$.

We define the Fourier transform over $r$ of $f_r(t,r)$ as:
\begin{eqnarray}
\tilde{f}_r(t,q)\equiv \int f_r(t,r) \sin(qr) dr\;\;.
\label{eq;rtok1}
\end{eqnarray}
As shown in Ref.\cite{Levashov20141}, $\tilde{f}_r(t,q)$ for non-decaying plane waves
should exhibit {\it constant amplitude} oscillations in $t$ with a period determined by the dispersion
relation.

Equation (\ref{eq;rtok1}) can also be rationalized from a different perspective.
It is natural to assume that the stress correlation function for a particular pair of atoms,
i.e., $f_p(t,\bm{r})$, depends on the direction of the radius vector, $\bm{r}$, 
from one atom to another. 
Let us define the three-dimensional Fourier transform of this stress correlation function 
as it is usually done: 
\begin{eqnarray}
\tilde{f}_p(t,\bm{q})\equiv \int f_p(t,\bm{r}) \exp(\;-i\bm{q}\bm{r}\;)d\bm{r}\;\;.
\label{eq;rtok2}
\end{eqnarray}
In isotropic cases, $f_p(t,\bm{r})\equiv f_p(t,r)$ and 
(\ref{eq;rtok2}) could be rewritten as:
\begin{eqnarray}
\left[q\,\tilde{f}_p(t,q)\right] \equiv \int \left[r\,f_p(t,r)\right] \sin(qr) dr\;\;.
\label{eq;rtok3}
\end{eqnarray}
It follows from (\ref{eq;fpandfr}) that expression (\ref{eq;rtok3}) is equivalent to
expression (\ref{eq;rtok1}).
The expression (\ref{eq;rtok3}) is similar to the expression that 
connects the pair distribution
function, to the reduced scattering intensity \cite{Warren}.

The Fourier transform in time-space naturally follows from the formulas (\ref{eq;ttow},\ref{eq;rtok1}).
It was shown in Ref.\cite{Levashov20141}, in the frame of the model considered there, 
that the Fourier transform of $f_r(t,r)$ over $t$ and $r$ should lead to the dispersion curves.

\section{Results of the Fourier Transforms at 1500 K \label{s:results1500}}

Since $F(t,r)$ was obtained in MD simulations on systems of finite sizes with periodic 
boundary conditions there is a lower limit on the possible values of $q$ that we can consider. 
See Ref.\cite{qmin} for details.

\subsection{Time to frequency Fourier transform}

The second row of Fig.\ref{fig:Fig1} shows $\tilde{F}(\omega,r)$, i.e., time to frequency 
Fourier transforms (\ref{eq;ttow}) of $F(t,r)$ and contributions to it from 
the \wave-like and the \pdf-like parts. 
Panels (2,1), (2,2), (2,3) were obtained from the data in panels
(1,1), (1,2), (1,3) respectively. 

\begin{figure}
\begin{center}
\includegraphics[angle=0,width=3.3in]{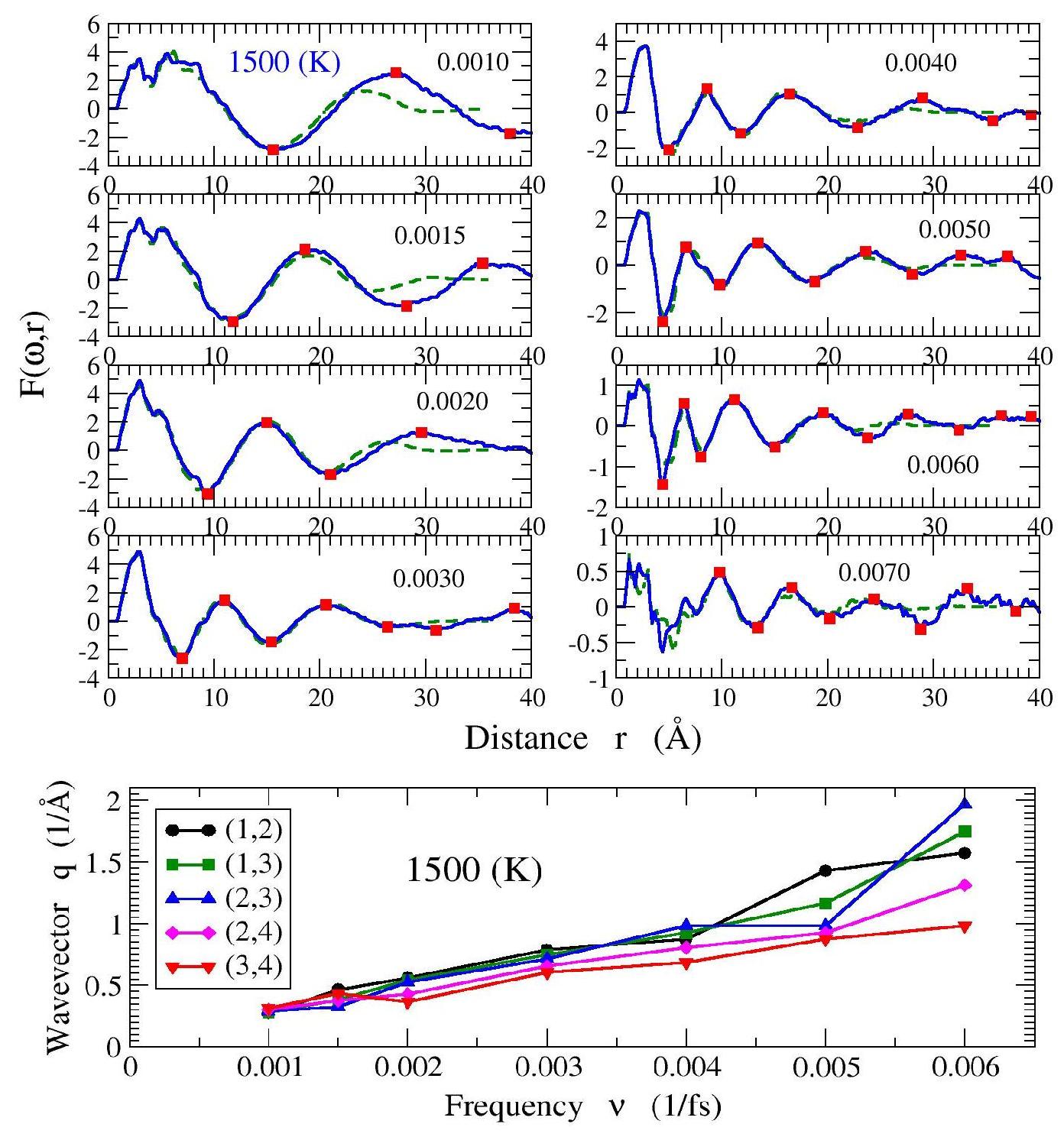}
\caption{Constant $\omega$ cuts from panel (2,2). 
Legends show the values of $\omega$.
The blue curves in upper two panels show the result from the large system with $(L/2) \approx 41.2$ (\AA).
Green curves show the results from the system with $(L/2) \approx 20.6$ (\AA).
We use the positions of the maximums and minimums, marked by the red squares, to determine 
the wavelengths, $\lambda = 2\pi/q$. Lower horizontal panel shows the dependence
of $q$ on $\omega$. Different curves in the lower panel correspond to different
selections of red squares used for the determination of wavelength. For example,
the notation (2,4) corresponds to the selection of the second and the forth
squares from the left to determine the wavelength. 
}\label{fig:Fig05}
\end{center}
\end{figure}

In panel (2,2) contributions from the shear and compression waves
are mixed. For an analysis of the stress waves it would be very
useful to find a way to separate contributions from these
waves. Since the amplitude of the compression wave in panel (1,2)
is significantly smaller than the amplitude of the shear wave,
it is reasonable to assume that features in the upper panels of Fig.\ref{fig:Fig05}
are dominated by the shear waves. 

It is useful to compare panel (2,2) of Fig.\ref{fig:Fig1} of this paper with panel (1,2) of Fig.7 in Ref.\cite{Levashov20141}.
Note, however, that panel (2,2) of Fig.\ref{fig:Fig1} shows $t$ to $\omega$ Fourier transform
of the function $r^2 f_p(t,r)$, while panel (1,2) of Fig.7 in Ref.\cite{Levashov20141}
shows the Fourier transform of the function $r f_p(t,r)$. We show
in this paper the Fourier transform of  $r^2 f_p(t,r)$ because it is more directly related to the generalized viscosity
and also because in Fig.\ref{fig:Fig05} this $r^2$-scaling allows showing relative amplitudes of the peaks
in $\tilde{F}(\omega,r)$ more clearly.

Figure \ref{fig:Fig05} shows constant $\omega$-cuts of panel
(2,2). 
If in panel (2,2) there were only shear waves, then, 
according to Ref.\cite{Levashov20141},
for every $\omega$ in Fig.\ref{fig:Fig05} the period of oscillations 
in $r$ would give the wavelength 
that corresponds to this value of $\omega$.
If  the \sscf were caused by non-decaying plane waves, then
the amplitudes of the peaks in  Fig.\ref{fig:Fig05} would linearly increase
with increase of $r$. 
However, the amplitudes of the peaks in Fig.\ref{fig:Fig05} decrease with
increase of $r$. This behavior suggests that the dynamic
underlying the behavior of $F(t,r)$ is very different from 
the vibrational dynamics of non-decaying plane waves.

The lower panel in Fig.\ref{fig:Fig05} shows the dependence of wavevector 
on frequency determined from the two upper panels.
This dependence should primarily correspond to the dispersion
relation for the shear waves. Indeed, the slope of the curves corresponds
to the speed $\approx 3$ (km/s), i.e., to the shear waves, according to panel (2,2) 
of Fig.\ref{fig:Fig1}.
Still, this picture should  contain certain distortions due to the compression waves.

By comparing the scales on the $y$-axes in the upper 
panels of Figure \ref{fig:Fig05}, note that the amplitudes of the waves for higher frequencies
are significantly smaller than the amplitudes for lower frequencies.
Note also that, even for the highest frequencies shown,
the waves propagate over, at least, 5 interatomic spacings ($r_{nn} \approx 2.6$ (\AA)).
In this context the following comment is relevant.
It can be seen in panel (1,2) of Fig.\ref{fig:Fig1}
and in Fig.\ref{fig:Fig03} that the feature corresponding 
to the compression waves is not just smaller in the amplitude
than the feature corresponding to the shear waves, but it is also narrower in $t$ and in $r$. 
Since it is narrower in $t$ its Fourier transform over time decays in a wider range of frequencies.
Thus contributions from the compression waves to the higher frequency curves in Fig.\ref{fig:Fig05} 
should be relatively larger than to the lower frequency curves. 
Because of the overlap of contributions from the compression and shear waves,
we do not discuss here attenuation rates for different frequencies. 

In considerations of the {\it macroscopic} (\tccf) it is assumed that only transverse waves contribute to it 
\cite{HansenJP20061,Boon19911,EvansDJ19901,Levesque19731,EvansDJ19811,Alder19831,
Mountain19821,PalmerBJ19941,Kaufman2007E,Puscasu110,Tanaka20091,Tanaka2011E,
Furukawa2013E,Mizuno2012,Mizuno2013}.
However, in view of the results discussed above, it is likely
that compression waves also affect the \tccf.
Thus the results obtained from the analysis of the \tccf can be distorted by the
compression waves. While the distortions should not be very significant
this issue deserves attention and clarification.

It follows from panel (2,2) of Fig.\ref{fig:Fig1} that the main \sickle feature vanishes 
at large distances because of the finite system size. 
This effect can also be seen in Fig. \ref{fig:Fig09}.
Thus, periodic boundary conditions (\pbc) affect the stress waves of small frequencies, 
i.e., $\nu \approx 0.0005-0.001$ (fs\textsuperscript{-1}). 
It is shown in section (\ref{s:tongue}) that contributions from the shear and compression waves 
overlap in the main \sickle feature.

Flattening of the main \sickle feature at low temperatures 
in the region of frequencies between 0.001 and 0.002 (fs\textsuperscript{-1})
means that the stress waves of the lower frequencies can propagate 
over  large distances. 
Frequency $\nu =0.001$ (fs\textsuperscript{-1}) corresponds to the energy $h\nu\approx 4.1$ (meV).
This energy approximately corresponds to the energy of the boson peak in 
metallic glasses \cite{Ruocco20131,Ediger20121}.
Thus significant increase of the propagation range with decrease of temperature
happens in the range of frequencies usually associated with the boson peak.
In a recent review \cite{Ediger20121} it was stated, on the basis 
of Ref.\cite{Taraskin20021,Tanaka20081}, that:
``There appears to be a growing consensus that the frequency of the boson peak
corresponds to the maximum frequency at which transverse phonons can 
propagate in the disordered material ..."
Our data are in agreement with this statement.

In panel (2,3) seemingly faster decay of the vertical stripes at large distances is misleading.
Perceived behavior originates simply from the smaller amplitudes of the stripes at large distances at zero time.
According to  Fig.\ref{fig:Fig04} at large distances all stripes decay at the same rate.

\section{Frequency dependent viscosity \label{s:etaw}}

According to formulas (\ref{eq;etakw-2}-\ref{eq;etakw-5})
in the Appendix  \ref{ax:tccf} viscosity is a complex function of the wavevector $\bm{k}$ and frequency $\omega$.
The stress correlation function is a complex function of $\bm{k}$ and time.
All results presented in this paper have been obtained for $\bm{k}=0$. 
In this case the components of the stress tensor and their correlation functions are real quantities 
(\ref{eq;etakw-4a},\ref{eq;etakw-4},\ref{eq;etakw-5},\ref{eq;etakw-8}). 
However, viscosity remains a complex function of $\omega$ (\ref{eq;etakw-7}):
$\eta(\omega) \equiv \eta'(\omega) - i\eta''(\omega)$. 
Complex viscosity is related to the complex shear modulus: 
$G(\omega) =G'(\omega) + i G''(\omega)= i\omega \eta(\omega)$ \cite{Bland}.
Thus the real part of viscosity describes energy dissipation in liquids, 
while the imaginary part describes elastic response.

It follows from the previous definitions of  $\eta(\omega)$ \cite{HansenJP20061,Boon19911,EvansDJ19901,Kaufman2007E}
and our definitions \cite{Levashov2013,Levashov20111}
 that:
\begin{eqnarray}
\eta'(\omega) = \frac{\rho_o}{k_b T} \int_{0}^{t_{max}} \left\{ \int_{0}^{R_{max}}  
F(t,r) dr \right\}\cos(\omega t)\,dt\;\;\;.
\label{eq;cossin}
\end{eqnarray}
Or:
 \begin{eqnarray}
\eta'(\omega) = \frac{\rho_o}{k_b T} \int_{0}^{R_{max}} \tilde{F}(\omega,r) dr\;\;\;.
\end{eqnarray}
Thus, for every $\omega$ in panel (2,1) of Fig.\ref{fig:Fig1} the integral over $r$
gives $\eta'(\omega)$. 
Integration over a range of distances, $(r_1,r_2)$,  should allow estimation of 
how this range contributes to $\eta'(\omega)$. 

\begin{figure}
\begin{center}
\includegraphics[angle=0,width=3.0in]{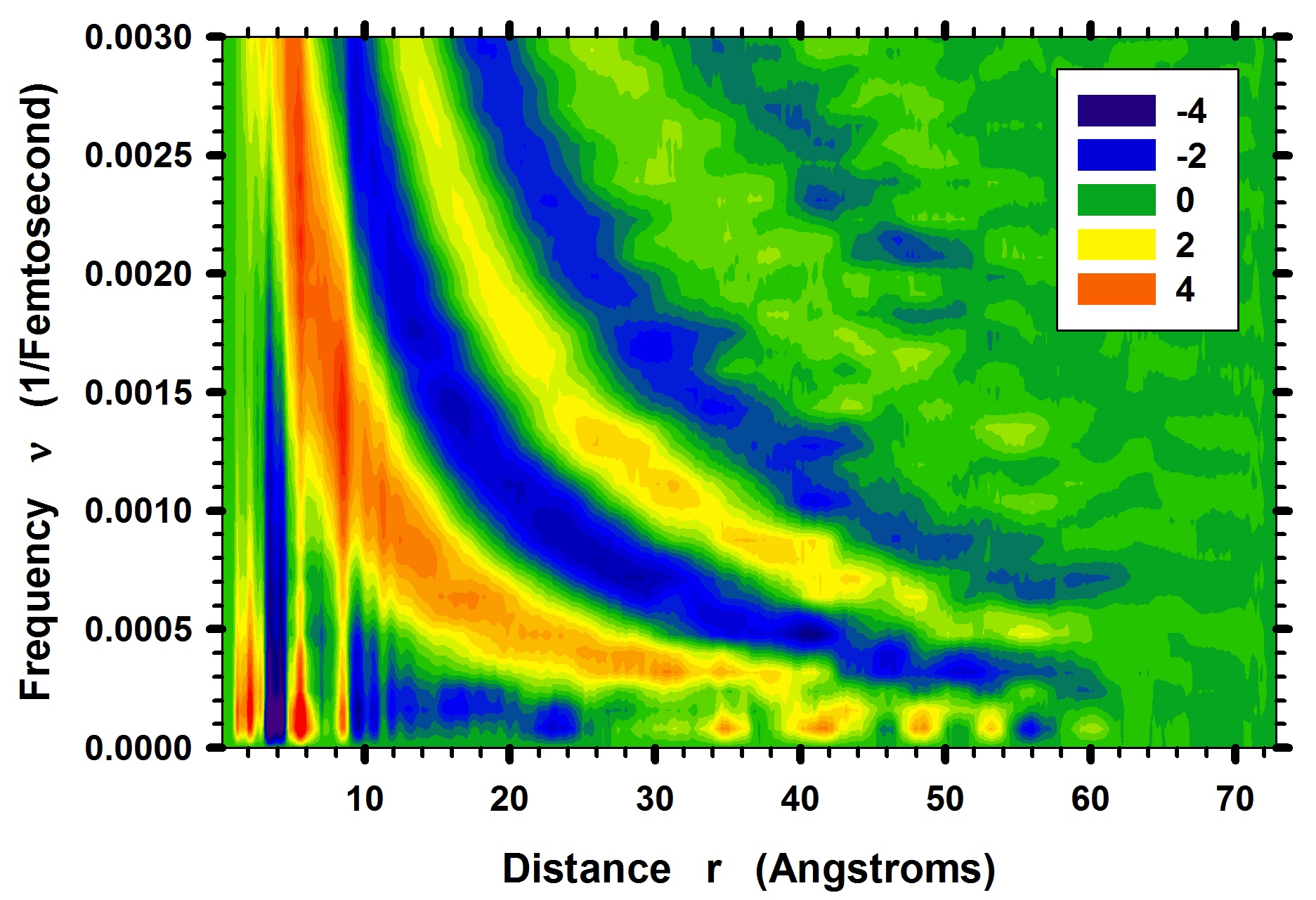}
\caption{Fourier transform of the data in panel (1,2) of Fig.\ref{fig:Fig1} through
$\sin(\omega t)$. This figure is very similar to panel (2,2) of
Fig.\ref{fig:Fig1}. However, it is a different figure.
For every $\omega$ the integral over $r$
gives the imaginary part of viscosity, i.e., $\eta''(\omega)$. 
The largest time up to which we calculated the \sscf in our simulations was 4000 (fs).
Thus we should not consider frequencies smaller than $1/4000$ (1/fs) or $0.00025$ (1/fs).
}\label{fig:etapp2d}
\end{center}
\end{figure}

According to formula (\ref{eq;etakw-7}) the imaginary part of viscosity can be calculated with
$\sin(\omega t)$ instead of $\cos(\omega t)$ in (\ref{eq;cossin}). 
Figure \ref{fig:etapp2d} shows $\sin(\omega t)$ Fourier transform of the panel (1,2) of Fig.\ref{fig:Fig1} 
in the region of smaller frequencies. 
For every $\omega$ in Fig.\ref{fig:etapp2d} the integral over $r$
gives the imaginary part of viscosity, i.e., $\eta''(\omega)$. 
We will see further that the most interesting behavior happens in the region $\omega < 0.0010$ (1/fs).

It is known that the real part of the frequency dependent viscosity, $\eta'(\omega)$, exhibits on
decrease of temperature frequency dependent increase \cite{Alder19831,Kaufman2007E,Tanaka20091,Donko20101}. 
This increase is the most significant for small $\omega$.
Considerations of our results in this context provide additional insights into this phenomena.

\begin{figure}
\begin{center}
\includegraphics[angle=0,width=3.3in]{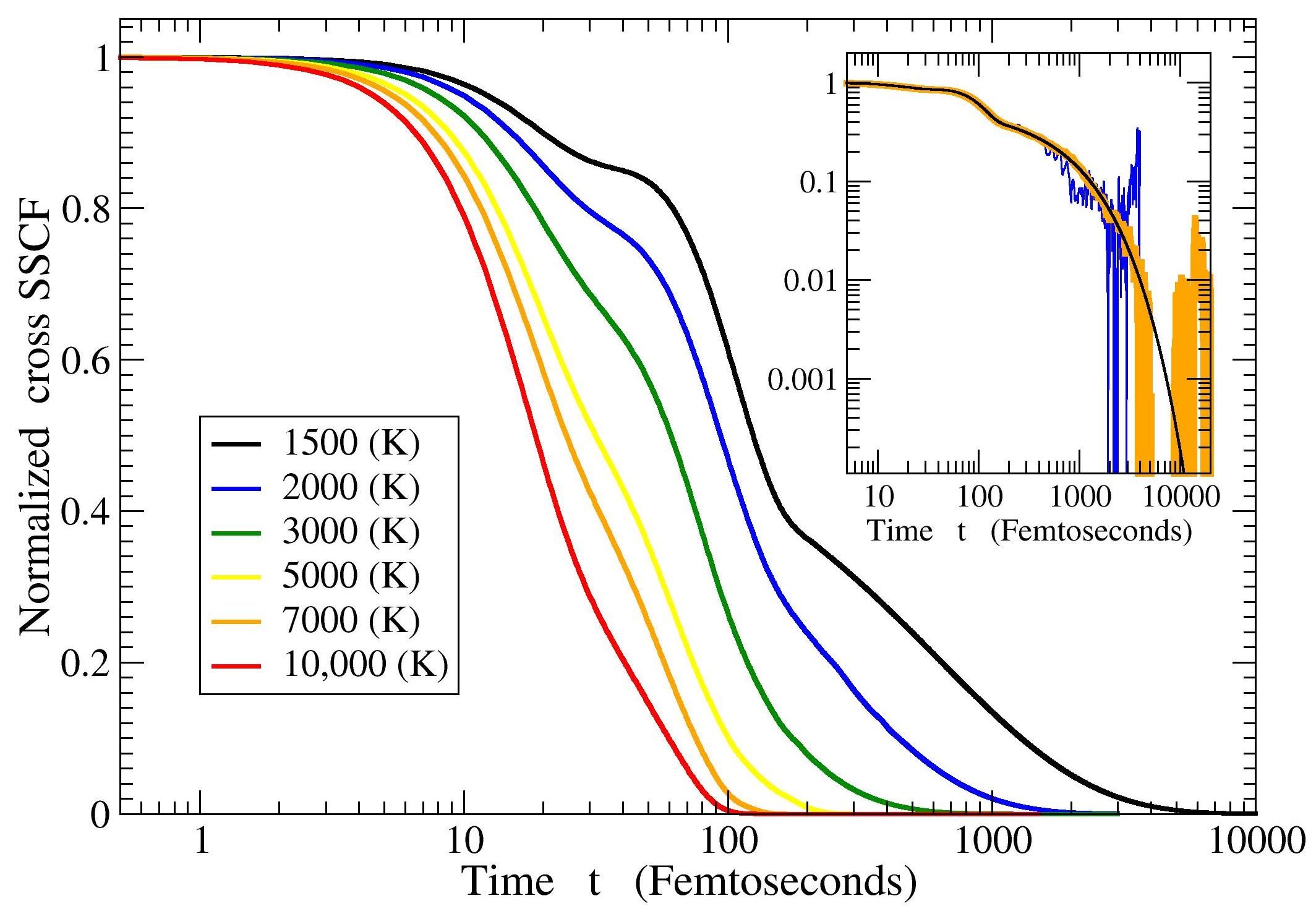}
\caption{Normalized \cross \sscfs at different 
temperatures. In the inset the blue curve shows the results of integration
over $r$ of the data in panel (1,1) of Fig.\ref{fig:Fig1}. The orange
curve in the inset was obtained on the intermediate size system as a difference between the \total
\sscf and the \self term of the \sscf \cite{Levashov2013}.
The black curve at short times is the orange curve.
For times $t>230$ (fs) the black curve was fitted to the reliable part of the simulation data and then the fitted curve
was used as an approximation when the simulation data become unreliable \cite{fitdiff}.
The curves in the main plot show the results of MD simulations 
at short times and the fitted curves at large times \cite{fitdiff}.
}\label{fig:sscf-diff}
\end{center}
\end{figure}

The statistics of our data for the atomic level \sscf is not sufficient to consider in detail low
frequency behavior of the macroscopic viscosity due to the \cross term. 
However, the macroscopic \sscf due to the \cross term can also be obtained as a difference between the \total
\sscf function and the \self term of the \sscf.
In Ref.\cite{Levashov2013} we considered the behaviors of the \total \sscfs and their \self terms at different temperatures.
The differences between the \total \sscfs and their \self terms are shown 
in Fig.\ref{fig:sscf-diff}.

\begin{figure}
\begin{center}
\includegraphics[angle=0,width=3.0in]{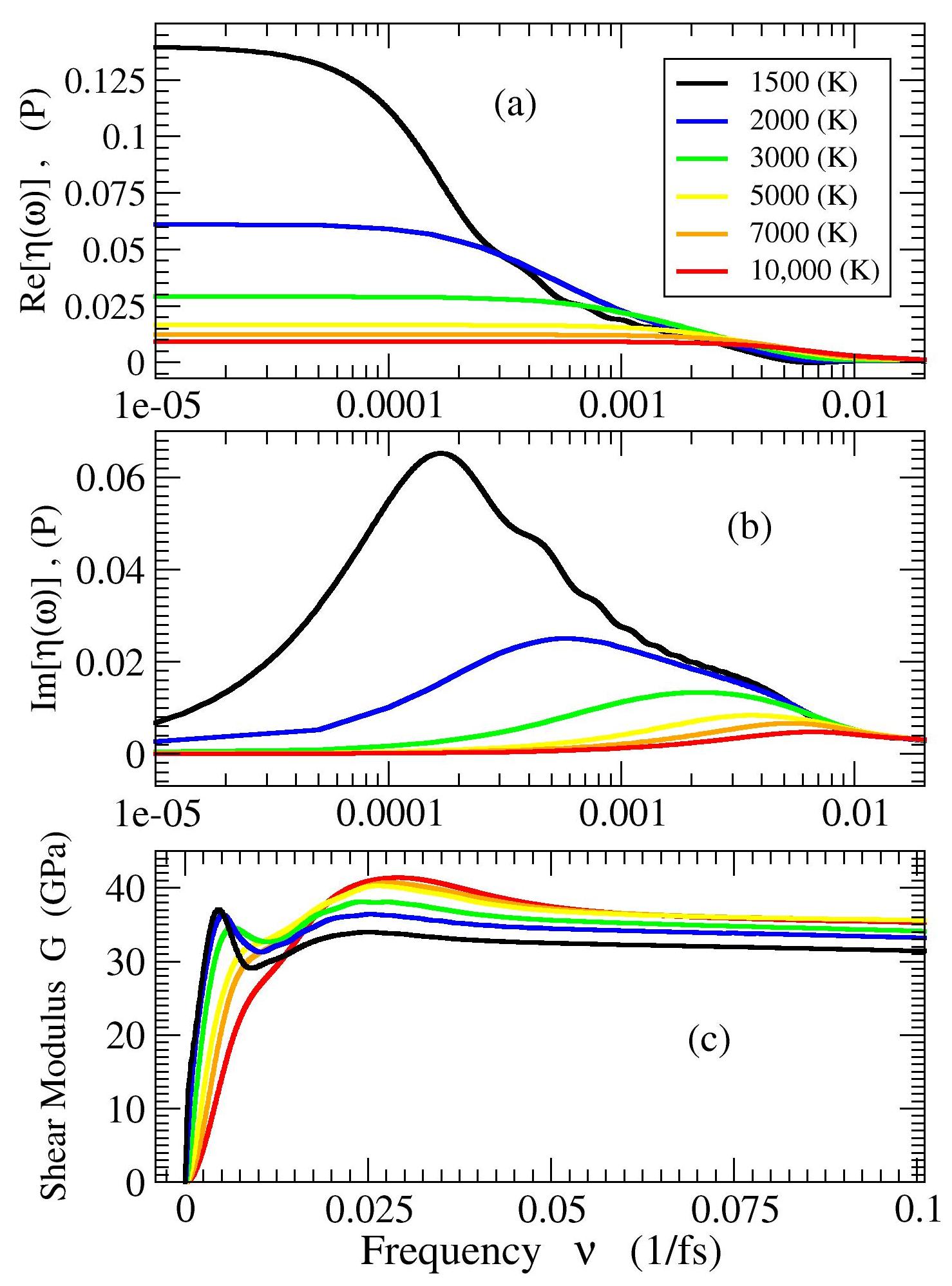}
\caption{Panel (a) shows the dependencies of $\eta'(\omega)$ on $\omega$ at 
different temperatures. The curves were obtained through $\cos(\omega t)$ integration 
of the curves in Fig.\ref{fig:sscf-diff}. 
Note that increase in the value of $\eta'(\omega)$ on decrease of temperature is the most significant for
low frequencies. This increase correlates with increase in the ranges of propagation of the shear waves 
of low frequencies on decrease of temperature.
Panel (b) shows dependencies of $\eta''(\omega)$ on $\omega$.
Panel (c) shows dependencies of the real part of the shear modulus, i.e. $\omega \eta''(\omega)$,
on $\omega$ at different temperatures.
Note that the scale on $\omega$-axis in (c) is different from $\omega$-scales in (a) and (b).
}\label{fig:eta-Re-Im}
\end{center}
\end{figure}

Panels (a) and (b) of Fig.\ref{fig:eta-Re-Im} show how the real and imaginary parts of viscosity due to the \cross term
depend on frequency. The curves in these panels were obtained by $\cos(\omega t)$ and $\sin(\omega t)$ integrations
of the \cross \sscf curves in Fig.\ref{fig:sscf-diff}. 
The viscosity curves exhibit expected behaviors. 
The rise in the value of the real part of viscosity at low frequencies was reported many times previously 
\cite{Alder19831,Kaufman2007E,Tanaka20091,Donko20101}.
The presence of the peak in the imaginary part of viscosity is also well known \cite{Bland,Donko20121,Maggi2008}.

Panel (c) shows the real part of the frequency dependent shear modulus, i.e., $G'(\omega)=\omega \eta''(\omega)$. 
In the limit of large frequencies the curves exhibit expected saturation to the infinite frequency value.
Infinite frequency shear modulus for {\it fcc} iron is $\sim 80$ (GPa) \cite{EightyGPa}. 
From Fig.\ref{fig:eta-Re-Im} we get the value $\sim 30$ (GPa).
This happens because we consider the contribution from the \cross term only.
The value of the shear modulus increases with increase of temperature because the data has been 
obtained in constant volume simulations.

New insights come from the comparisons of the regions in $\omega$ where $\eta'(w)$ and $\eta''(\omega)$
start to increase from their large-$\omega$ values with the corresponding $\omega$-regions
in panel (2,2) of Fig.\ref{fig:Fig1} and in Fig.\ref{fig:etapp2d}. 
These comparisons suggest that the increase in the ranges of propagation of the shear stress waves
correlates with the increase in the values of the real and imaginary parts of viscosity.
For $\eta''(\omega)$ the increase is related to the  sickle feature in Fig.\ref{fig:etapp2d} 
which is the closest to the origin.
We again note that we should not consider frequencies $\omega < 0.00025$ (1/fs) on the basis
of Fig.\ref{fig:Fig1} and Fig.\ref{fig:etapp2d} since corresponding simulations were not long enough.

\begin{figure}
\begin{center}
\includegraphics[angle=0,width=3.4in]{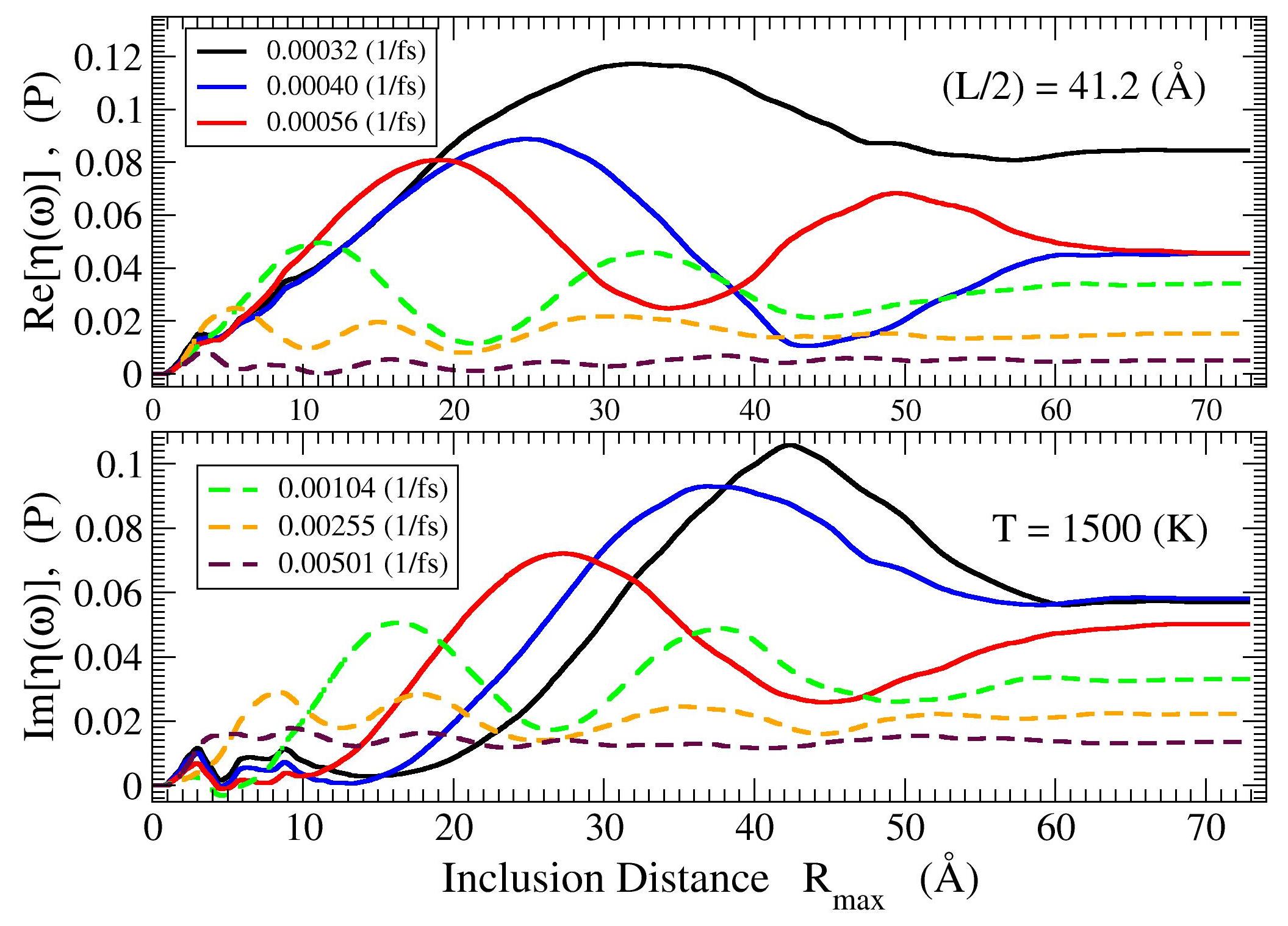}
\caption{Dependencies of $\eta'(\omega,R_{max})$ and $\eta''(\omega,R_{max})$ on the inclusion distance
$R_{max}$. The curves for $\eta'(\omega,R_{max})$ in panel (a) were obtained by integration over $r$ of the constant 
$\omega$-cuts from panel (2,2) of Fig.\ref{fig:Fig1}. The curves for $\eta''(\omega,R_{max})$ in panel (b) were obtained
by integration over $r$ of the constant $\omega$-cuts from Fig.\ref{fig:etapp2d}. 
The {\it macroscopic} values of the viscosities correspond to the values to which the curves converge at large $R_{max}$.
The selected frequencies are the same in both panels.
The legends in panel (a) give the values of $\omega$ for the solid curves. The legends in panel (b) give the values 
of $\omega$ for the dashed curves. Note that all solid curves correspond to the values of $\omega<0.001$ (1/fs).
}\label{fig:eta-sun-nu-r-1c}
\end{center}
\end{figure}

In order to demonstrate the connection between the propagation of the shear waves and viscosity further
we show in panel (a) of Fig.\ref{fig:eta-sun-nu-r-1c} how
$\eta'(\omega,R_{max})\equiv \int_0^{R_{max}} \tilde{F}(\omega,r)dr$ depends on $R_{max}$ for the selected values of 
$\omega$ shown in the legends. Panel (b) is similar to panel (a), but it is for $\eta''(\omega,R_{max})$. 
The shapes of the curves suggest/demonstrate that the microscopic origin of viscosity is related 
to the propagation and dissipation of the shear waves on atomic scale.

\subsection{Distance to wavevector Fourier transform}

Panels (3,1), (3,2), (3,3) show $r$ to $q$ Fourier transforms (\ref{eq;rtok1})
of the function $f_r(t,r)=rf_p(t,r)$ obtained from the data 
in panels (1,1), (1,2), (1,3) respectively.
It is useful to compare panel (3,2) with panel (2,1) of Fig.7 in Ref.\cite{Levashov20141}.
In these two panels $r$-scalings are the same.

For every particular time, the Fourier transform (\ref{eq;rtok1}) 
of the \sscf over $r$ is similar to the transform of 
the pair density function, $G(r)$, into the structure factor, $S(q)$; if 
$G(r)=4\pi r \left[ \rho(r)-\rho_o \right]$ then 
$q\left[S(q) - 1\right] = \int_{0}^{R_{max}}G(r)\,\sin(qr)\,dr$ \cite{Warren}.
Thus, knowledge of the general relations between $G(r)$ and $S(q)$ can help
in guessing the roles of certain features. 
This parallel allows us to relate the negative intensity
near 3 (\AA\textsuperscript{-1}) in (3,3) to the periodicity in $r$ of the \pdf-like
contribution to the \sscf (see (1,3)). The width of the 3 (\AA\textsuperscript{-1}) feature in (3,3)
is related to the extend of the \pdf-like oscillations in $r$.

Panels (3,1) and (3,2) show for how long in time stress waves, 
with a particular value of the wavevector, exist. 
If the stress waves were non-decaying plane waves, then,
according to Ref. \cite{Levashov20141},
for every $q$ the amplitude of oscillations would be constant in time.
Note that non-zero intensity for smaller $q$ exists for larger times than nonzero intensity for larger
wavevectors. 
Recall that in panels (3,1) and (3,2) the contributions from 
the shear and compression waves overlap.
 
\begin{figure}
\begin{center}
\includegraphics[angle=0,width=3.3in]{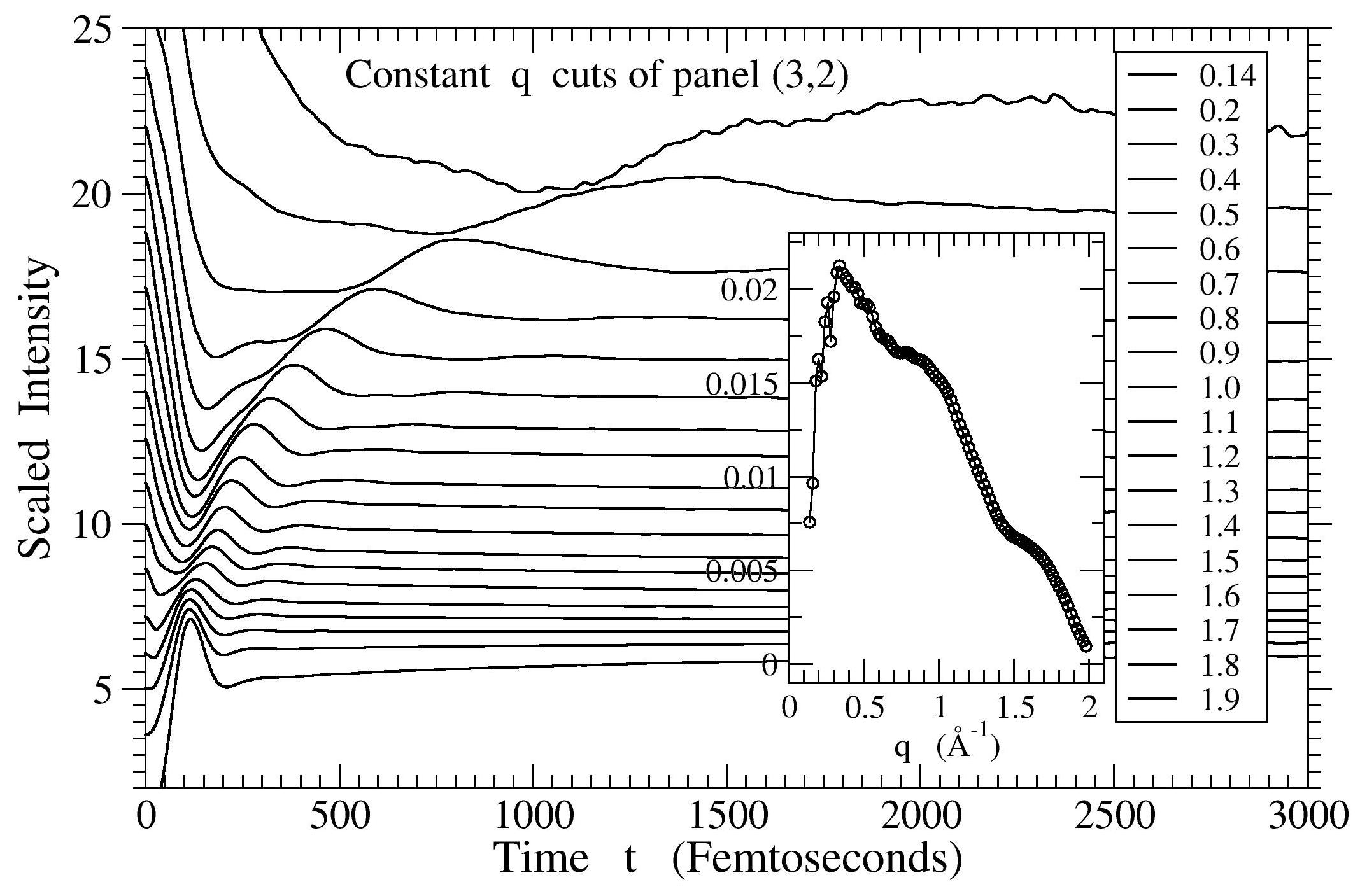}
\caption{
Normalized and shifted constant $q$ cuts of panel (3,2) of Fig.\ref{fig:Fig1}. 
Every particular $q$-cut was scaled to the maximum intensity in the \sickle feature for this
$q$ and then shifted. The dependence of this maximum intensity on $q$ is shown in the inset. 
Different curves correspond to the different values of $q$ (in \AA\textsuperscript{-1}) given in the legends. 
Upper curves correspond to the upper legends.   
}\label{fig:Fig07}
\end{center}
\end{figure}

Figure \ref{fig:Fig07} shows constant $q$ 
cuts of panel (3,2) of Fig.\ref{fig:Fig1}. 
We see in Fig.\ref{fig:Fig07}, as in panel (3,2), 
that for every $q$ there are no more than two oscillations in time, with
the amplitude of the second maximum significantly smaller than the amplitude of the 
first maximum.
Thus, the situation in the considered liquid at 1500 (K) is quite different
from the situation in a model crystal with non-decaying vibrational plane waves \cite{Levashov20141}. 
In the model crystal the amplitudes of these oscillations should be constant in time.
The relative intensities of the maximums in Fig.\ref{fig:Fig07} for 
different $q$ should be related to the relative vibrational densities of states and to the 
relative rates of decay for different $q$.

\begin{figure*}
\begin{center}
\includegraphics[angle=0,width=7.0in]{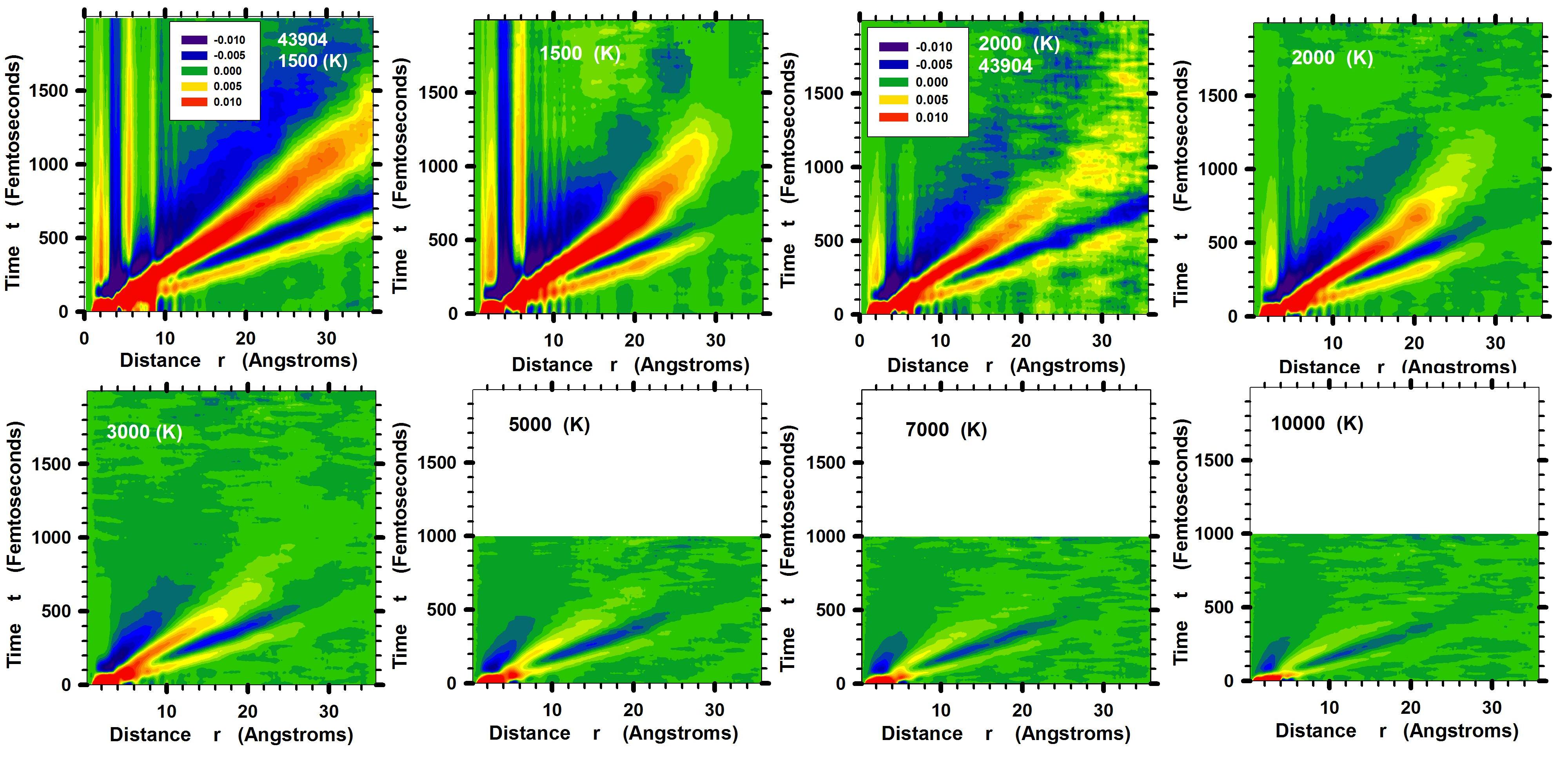}
\caption{The \wave-like contributions to the \sscf at different temperatures on the systems
of 43904 and 5488 particles.  Panels (1,1) and (1,3) are for the 43904 particle system.
All other panels are for the 5488 particle system. 
Panels (1,1) and (1,2) are for $T=1500$ (K). Panels (1,3) and (1,4) are for $T=2000$ (K).
Panel (2,1) is for 3000 (K), (2,2) is for 5000 (K), (2,3) is for 7000 (K), (2,4) is for 10000 (K).
Note in panels (1,1) and (1,2) bright vertical stripes at $r \sim 5$ (\AA). These stripes
show that the separation procedure used to produce \wave-like contributions does not work
for these distances. Note also that there are no such bright vertical stripes in the other panels.
In the text we argue that this bright vertical stripe is the origin of the \bonfire feature in
the (t,q)-\sscf.
}\label{fig:Fig08}
\end{center}
\end{figure*}

Note that the main \sickle feature ends at $q_{wmax} \approx 1.75$ (\AA\textsuperscript{-1}), i.e., at
 $\lambda_{wmin} =2\pi/q_{wmax} \approx 3.6$ (\AA).
 Thus, $\lambda_{wmin} \approx 1.4d$, where $d$ is the average distance between the nearest particles for the chosen
 value of the density and also the equilibrium distances between a pair of particles for our potential. 
The fact that the smallest possible wavelength of the shear stress waves 
is $\approx 1.5\,d$ is in approximate agreement with the other results
\cite{Alley19831,Mizuno2013}. 
In our data, the crossover in the main \sickle feature happens at $q_c \approx 0.5$ (\AA\textsuperscript{-1}), i.e., 
at $\lambda_c \approx 12.6$ (\AA) or $\approx 4.8d$. At larger distances,
according to Ref.\cite{Mizuno2013}, {\it ordinary} hydrodynamics with $q$-independent 
transport coefficients is valid, while at smaller distances the situation
is more complicated.

\begin{figure*}
\begin{center}
\includegraphics[angle=0,width=7.0in]{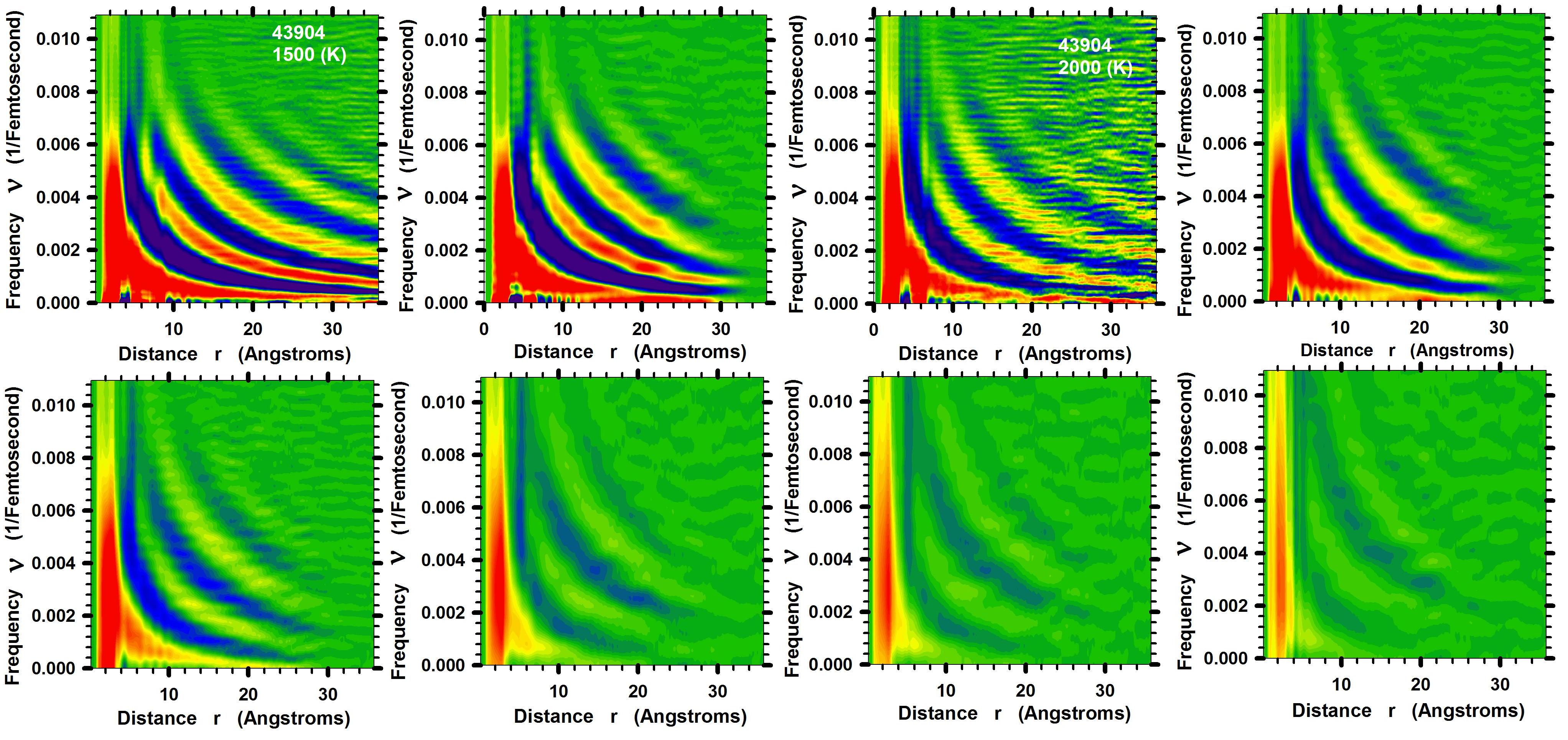}
\caption{The \sscf  in $(\omega,r)$-space for different systems and at different temperatures. 
The locations of panels correspond to those in Fig.\ref{fig:Fig08}. 
The scales on the $z$-axes are the same as in panel (2,2) of Fig.\ref{fig:Fig1}.
}\label{fig:Fig09}
\end{center}
\end{figure*}

\begin{figure*}
\begin{center}
\includegraphics[angle=0,width=7.0in]{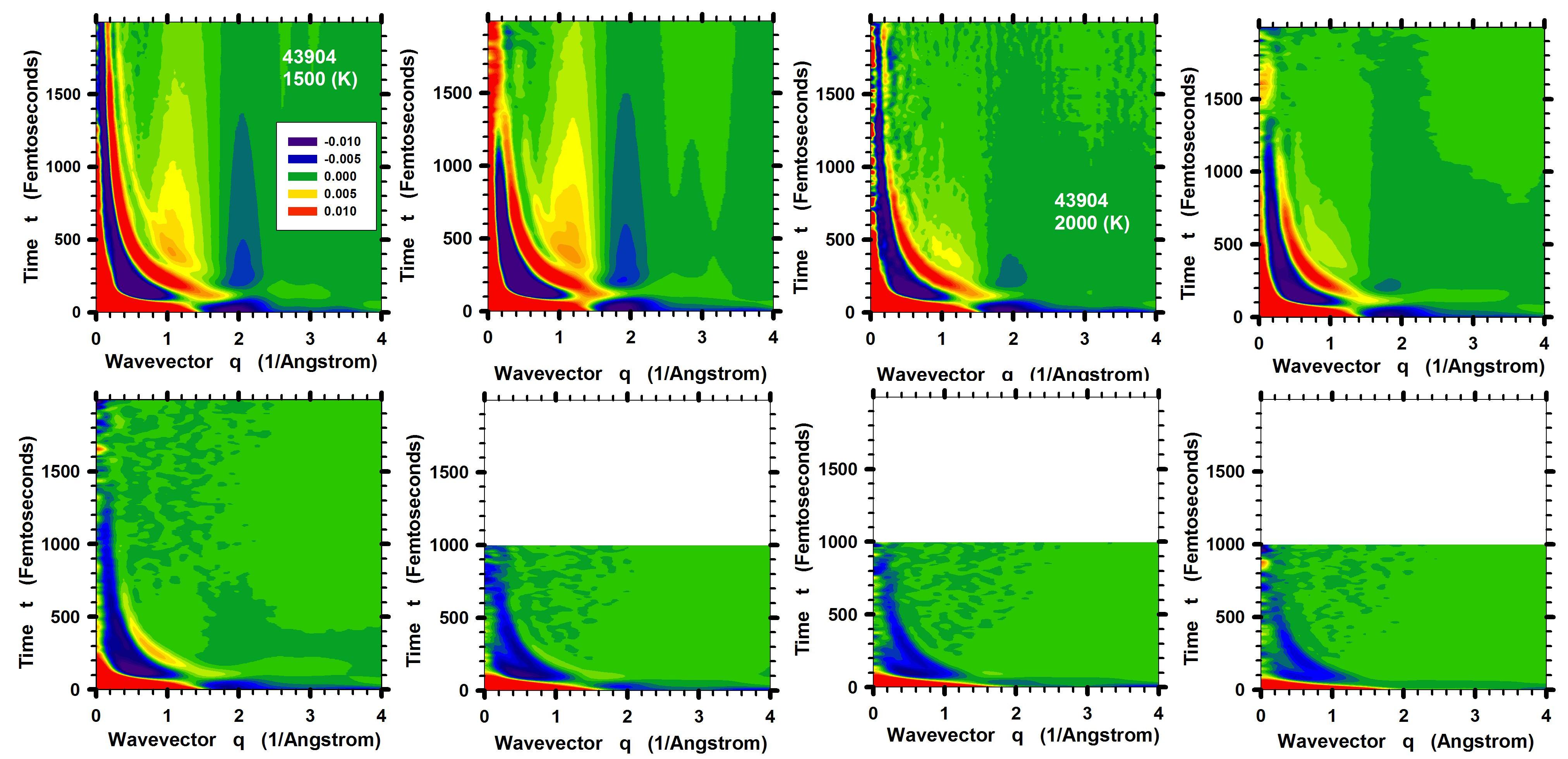}
\caption{The \sscf  in $(t,q)$-space for different systems and at different temperatures. 
The locations of panels correspond to those in Fig.\ref{fig:Fig08}.
Note that panels (1,1) and (1,2) contain the \bonfire feature, while
the other panels do not. From comparisons with panels
in Fig.\ref{fig:Fig08}, it follows that the \bonfire
feature is related to the bright vertical stripe at $r \approx 5$ (\AA).
}\label{fig:Fig10}
\end{center}
\end{figure*}

There are two features in panels (3,1) and (3,2) that we discuss in sections  (\ref{s:bonfire}) and (\ref{s:tongue}). 
The first feature is a positive intensity that
is centered at $q\approx 1.15\;\AA^{-1}$ and 
extends in time from approximately 300 (fs) 
to 1700 (fs). We call this feature the \bonfire.
Another feature extends in time 
from 0 to 200 (fs) and in $q$ from 0 to 1.3 (\AA\textsuperscript{-1}). 
We call this feature the \tongue. 

It turns out, that both features originate from the interval of distances between 
$\approx 2$ (\AA) and $\approx 7.5$ (\AA).  
The shape of the \tongue feature is affected by the position of the origin of  the stress waves
(they start from the first coordination shell and not from $r=0$ (\AA)). 
The \bonfire feature is related to the famous splitting of the second peak in the pair distribution
function which is associated with some local arrangements of particles which 
agglomerate into larger domains \cite{Tomida19951,Stillinger19981,Malins2013}.
The \bonfire feature is also 
present in Fig.\ref{fig:Fig07}, 
though it is difficult to see it. 

It is also possible to consider, from panels (3,1) and (3,2), wavevector dependent 
viscosity and thus study how different times contribute to it. 
However, in view of the discussion
in Appendix (\ref{ax:tccf}), these considerations need more insights and we 
will not focus on them now. 

\subsection{Fourier transform in time and space}

Panels (4,1), (4,2), and (4,3) show the Fourier transforms in time and space of $f_r(t,r)$.
One can guess in (4,1) and (4,2) broad dispersion curves associated with the stress waves.
The dispersion, however, is not well pronounced.

\section{Evolution of the data with temperature \label{s:Tevol}}

In this section we address the evolution of the \sscf and its Fourier transforms with temperature.
We also discuss size effects by comparing the data on the intermediate system of 5488 particles 
with $(L/2) = 20.06$ (\AA), and on the large system of 43904 particles with $(L/2) = 41.21$ (\AA).

The \total \sscfs for the two systems at different temperatures are shown in Fig.4,5 
of Ref.\cite{Levashov2013}.
Figure \ref {fig:Fig08} of this paper shows \wave's contributions to the \sscfs for 
different temperatures and systems in $(t,r)$-space. 
The comparisons of panels (1,1) with (1,2) 
and (1,3) with (1,4) show size effects at low temperatures.
It is clear that the finite size of the system affects propagation of the stress waves.
It is also clear that the stress waves are more pronounced at 1500 (K) than at 2000 (K).
The results in the second row show gradual disappearance of the stress waves 
with increase of temperature.

Note in the results for 1500 (K) a bright vertical line at $r\approx 6$ (\AA).
Note also that there is not a well pronounced line in the results for 2000 (K).
Comparisons with the corresponding panels in Fig.\ref{fig:Fig10} suggest that this
vertical line is related to the \bonfire feature.

Figure \ref{fig:Fig09} shows $t$ to $\omega$ Fourier transforms of the \sscfs.
It follows from the comparisons of panels (1,1) with (1,2) and (1,3) with (1,4) that 
the main \sickle feature vanishes at large distances because of the finite system size.
In panel (2,2) of Fig.\ref{fig:Fig1}  the same \sickle feature extends to significantly
larger distances. Thus \pbc affect frequencies $\nu < 0.004$ (fs\textsuperscript{-1}). 

Figure \ref{fig:Fig10} shows the \sscfs in $(t,q)$-space.
Note that the Fourier transforms of the \sscfs in $(t,q)$-space obtained on the large and on the 
intermediate systems exhibit different behaviors at small $q$. 
These differences should be related to the finite size effects \cite{qmin}. 
See Fig.\ref{fig:Fig1} for a wider range in time.

\section{Conclusion \label{s:conclusion}}

We investigated the Fourier transforms of the atomic level 
Green-Kubo \sscfs obtained in MD simulations on a model liquid. 
These considerations demonstrate that
the atomic level \sscf can be used to study how lifetimes and ranges
of propagation of stress waves depend on their frequency and wavevector.
It was also demonstrated that the crossover from quasilocalized to propagating behavior 
occurs at the frequencies usually associated with the Boson peak, confirming previous 
results \cite{Taraskin20021,Tanaka20081,Ruocco20131,Ediger20121}.
We found that the ranges of propagation of the shear stress waves 
for frequencies less than half of the Einstein frequency extend well beyond 
the nearest neighbor shell.

As temperature decreases the ranges of propagation of low frequency stress waves increase.
Our results show that this increase is correlated with the increase in the value of low frequency
viscosity. 
Thus, at $T>2T_A$, where $T_A$ is the potential energy landscape crossover temperature \cite{Levashov2008E},
stress waves of all frequencies decay on the length scales of 10 interatomic distances or less.
As temperature is lowered, the increase in the ranges of propagation for the lower frequency waves
is more significant than for the higher frequency waves. 
Thus our results suggest that being able to understand the structural origin of the increase in
ranges of propagation of low frequency shear stress waves might also help in understanding the nature of viscosity
increase on approach of the glass transition.

The conclusions to which we arrived using our new method are expected,
in view of other publications
\cite{Alder19831,PalmerBJ19941,Kaufman2007E,Puscasu110,Tanaka20091,Tanaka2011E,Furukawa2013E,Mizuno2012,Mizuno2013}. 
However, in our view, investigations with this method compliment 
the results obtained with other methods.

Our data show that viscosity is related to the propagating stress waves. 
On the other hand, it was argued recently that at low temperatures 
relaxation of the shear stresses should become 
activated \cite{Abraham20121,Buchenau20111}.
It is of interest to study if viscosity at lower temperatures decouples
from the shear stress waves, or if activated dynamics is causing decay of the stress waves, but
viscosity remains related to them.
For this it would be necessary to study a different system 
as the system that we studied crystallizes at relatively high temperatures.

Our results also suggest that the decay of the \pdf-like part of the atomic level
\sscf at large times is related to diffusion of particles.

The fact that we see compression waves in the shear stress correlation function 
should be related to the existence of correlations between the different 
components of the atomic level stresses on the same site \cite{Kust2003a,Kust2003b}.
   
Our method has important shortcomings.
For example, one would not suppose from our results, 
as they are the averages over many atoms and times,
about the presence of force chains 
\cite{Cates19981,OHern20011,Lacevic20011,Bi20011,Desmond20131} or
chain-like displacements \cite {Donati20131}.
It appears that the spherical averaging that we perform
also averages out the long range Eshelby field present 
in the system \cite{Picard20041,Shall20111,Lemaitr20131}.
It is unclear if it is possible to see dynamic heterogeneity \cite{Hetero20131}
with our method. These shortcomings, however,
are also present in the \tccf technique and in other approaches that rely on
considerations of macroscopic quantities. 

A separate question of interest is in what range of distances can the \sscf 
be modeled using viscoelastic approximations? 
The separation procedure that we used to extract the \wave-like and the \pdf-like 
contributions to the \sscf suggests that  continuous approximation may not work for distances 
smaller than 3 or 4 interatomic distances, but can work for larger ranges. 
This is in agreement with some other results \cite{Alley19831,Sokolov19921,Tomida19951,Mizuno2012,Mizuno2013}.

In references \cite{Schepper1987,Schepper1988,Mryglod1995,Mryglod19971,Mryglod19972,Mryglod19973,Mryglod20051,Bertolini20111}
an approach based on consideration of the generalized modes has been developed. 
It has been shown that it is sufficient to consider a relatively small number of the generalized 
modes in order to describe liquids' dynamics with rather good precision. 
It would be interesting to use the approach developed in this and two preceding 
papers \cite{Levashov20111,Levashov2013} in order to investigate the atomic scale nature of the generalized modes.

\section{Acknowledgments} 

We would like to thank T. Egami, V.N. Novikov, and K.A. Lokshin for useful discussions.

\appendix

\section{On the connection between the transverse current correlation approach and our considerations \label{ax:tccf}}

As we discuss in this paper the dependence of the \sscf on the wavevector $q$ it is important to note that the 
wavevector $\bm{q}$ that enters into our considerations is {\it distinct} 
from the wavevector $\bm{k}$ that usually enters into the discussions of the \tccf. 

In derivations of the expressions for generalized viscosity through correlation functions 
the wavevector $\bm{k}$ is related, in particular, to the density fluctuations. 
As we introduce the wavevector $q$, it is not formally related to the density fluctuations.
Standard considerations of the \tccf are as follows \cite{HansenJP20061,Boon19911,Tanaka2011E}. 

The transverse current, $\bm{j}^{\perp}_{\bm{k}}(t)$, and the transverse current correlation 
function, $C(k,t)$ are defined as:
\begin{eqnarray}
&&\bm{j}^{\perp}_{\bm{k}}(t) \equiv \frac{1}{N}\sum_{i=1}^{N} m_i \bm{v}^{\perp}_i(t)\exp\left(i\bm{k}\bm{r}_i(t)\right)\;,\;\;\;\\
&&C(t,k) \equiv \left< \bm{j}^{\perp}_{\bm{k}}(t) \bm{j}^{\perp}_{\bm{-k}}(0)\right>\;,\;\;\;\;\;\;\bm{v}^{\perp}_i \equiv \bm{v}_i - \bm{\hat{k}}\bm{v}_i\;\;.\;\;\;\;\;\;
\label{eq;jkt-1}
\end{eqnarray}

It was shown in the generalized hydrodynamics approach
that in an isotropic liquid the \tccf, $C(k,t)$, is associated with the wavevector and frequency 
dependent viscosity \cite{HansenJP20061,Boon19911}:
\begin{eqnarray}
&&\eta(\omega,k) = \frac{\rho_m}{k^2\widetilde{C}(\omega,k)}\left[-i\omega \widetilde{C}(\omega,k)+C(0,k)\right]\;,\;\;\;\\
&&\widetilde{C}(\omega,k) \equiv \int_0^{\infty}C(t,k)\exp(-i\omega t)dt\;\;,\;\;\;
\label{eq;etakw-1}
\end{eqnarray}
where $\rho_m = m N/V$ is the average mass density.
The {\it usual} viscosity corresponds to the limit of vanishing frequency and wavevector
 ($\omega \rightarrow 0,\;\;\bm{k} \rightarrow 0$). 
In this limit the following expression for viscosity in terms of the \tccf can be obtained:
\begin{eqnarray}
\eta=\beta \rho_m\lim_{\omega \to 0} \lim_{k \to 0}\textrm{Re} 
\int_0^{\infty}\frac{C(t,k)}{k^2}e^{-i\omega t} dt\;\;,\;\;\;\;\;\;\;
\label{eq;etakw-6}
\end{eqnarray}
where $\beta = (k_b T)^{-1}$.

Wavevector and frequency dependent viscosity 
can also be expressed through the correlation function of 
the macroscopic stress tensor, $\Pi_{\bm{k}}^{xz}(t)$ 
\cite{EvansDJ19811,EvansDJ19901,Alder19831,HansenJP20061}:
\begin{eqnarray}
\eta(\omega,k)= \left[\widetilde{N}(\omega,k)\right] / 
\left[1 - \left(k^2 \widetilde{N}(\omega,k)\right)/\left(i\omega \rho_m\right) \right]\;,\;\;\;\;\;\;\;\;
\label{eq;etakw-2}
\end{eqnarray} 
where 
\begin{eqnarray}
\widetilde{N}(\omega,k) = \left(\beta/V\right)\int_0^{\infty}
\left<\Pi_{\bm{k}}^{xz}(t)\Pi_{\bm{-k}}^{xz}(0)\right>e^{-i\omega t} dt\;\;\;\;\;\;\;\;\;\;\;\;
\label{eq;etakw-3}
\end{eqnarray}
and it is assumed that $\bm{k}$ is parallel to the z-axis.
The expression for the stress tensor in (\ref{eq;etakw-3}) is \cite{HansenJP20061}:
\begin{eqnarray}
\Pi_{\bm{k}}^{\alpha \beta}(t)=
\sum_{i=1}^{N} s^{\alpha \beta}_i(t,k)\;\;,
\label{eq;etakw-4a}
\end{eqnarray}
where  $s^{\alpha \beta}_i(t,k)$ is  the atomic level stress element:
\begin{eqnarray}
s^{\alpha \beta}_i(t) =
\left[m_i v_i^{\alpha}v_i^{\beta} +
\sum_{j \neq i}^{N}\frac{r_{ij}^{\alpha}r_{ij}^{\beta}}{r_{ij}^2}\Phi_{\bm{k}}(\bm{r}_{ij})
\right]e^{-i\bm{k}\bm{r}_{i}}\;\;\;\;\;\;\;\;\;
\label{eq;etakw-4}
\end{eqnarray}
and
\begin{eqnarray}
\Phi_{\bm{k}}(\bm{r}_{ij})=\frac{r_{ij}}{2}
\left[\frac{\partial U(r_{ij})}{\partial r_{ij}}\right]
\left[\frac{e^{i\bm{k}\bm{r}_{ij}} - 1}{i\bm{k}\bm{r}_{ij}}\right]\;\;.\;\;\;\;
\label{eq;etakw-5}    
\end{eqnarray}
In liquids at low temperatures the first term on the right hand side of (\ref{eq;etakw-4})
is much smaller than the second term and can be neglected  \cite{HansenJP20061}.

In the limit ($k \rightarrow 0$)  from (\ref{eq;etakw-2},\ref{eq;etakw-3}) we get: 
\begin{eqnarray}
\eta(\omega) = \frac{\beta}{V}
\int_{0}^{\infty}
<\Pi_{\bm{0}}^{xz}(t)\Pi_{\bm{0}}^{xz}(0)>
e^{-i\omega t} dt\;\;\;.\;\;\;\;\;\;\;
\label{eq;etakw-7}
\end{eqnarray}
For $\omega = 0$ expression (\ref{eq;etakw-7}) is the standard Green-Kubo formula.

It can be seen from (\ref{eq;etakw-3},\ref{eq;etakw-4a},\ref{eq;etakw-4},\ref{eq;etakw-5}), 
that $\widetilde{N}(k,\omega)$  can be decomposed into 
contributions from different atomic level stress elements.
In Ref.\cite{Levashov2013,Levashov20111} we studied the properties of this
decomposition for $\bm{k}=0$. Thus we introduced there a function $F(r,t)$:
\begin{eqnarray}
\left<\Pi_{\bm{0}}^{xz}(t)\Pi_{\bm{0}}^{xz}(0)\right> = \int_0^{\infty}F(r,t)dr \;\;.
\label{eq;etakw-8}
\end{eqnarray}
In this paper, we investigate features of $F(r,t)$ 
by performing the Fourier transform of $F(r,t)$ into 
$\widetilde{F}(q,t),\widetilde{F}(r,\omega),\widetilde{F}(q,\omega)$.

We would like to emphasize that the wavevector $\bm{q}$ that we introduce
in our present investigation {\it is distinct} from the wavevector $\bm{k}$ 
that is usually introduced in consideration of the generalized viscosity.
Formally all our results correspond to the case $\bm{k}=0$, 
i.e., to the case of very large wave lengths of density fluctuations.
This limit effectively corresponds to the case when 
local density fluctuations are absent.
 See also discussion on the transverse current correlation function 
in Ref.\cite{Levashov20141}.

\section{The origin of the {\it bonfire} feature \label{s:bonfire}}

\begin{figure}
\begin{center}
\includegraphics[angle=0,width=3.4in]{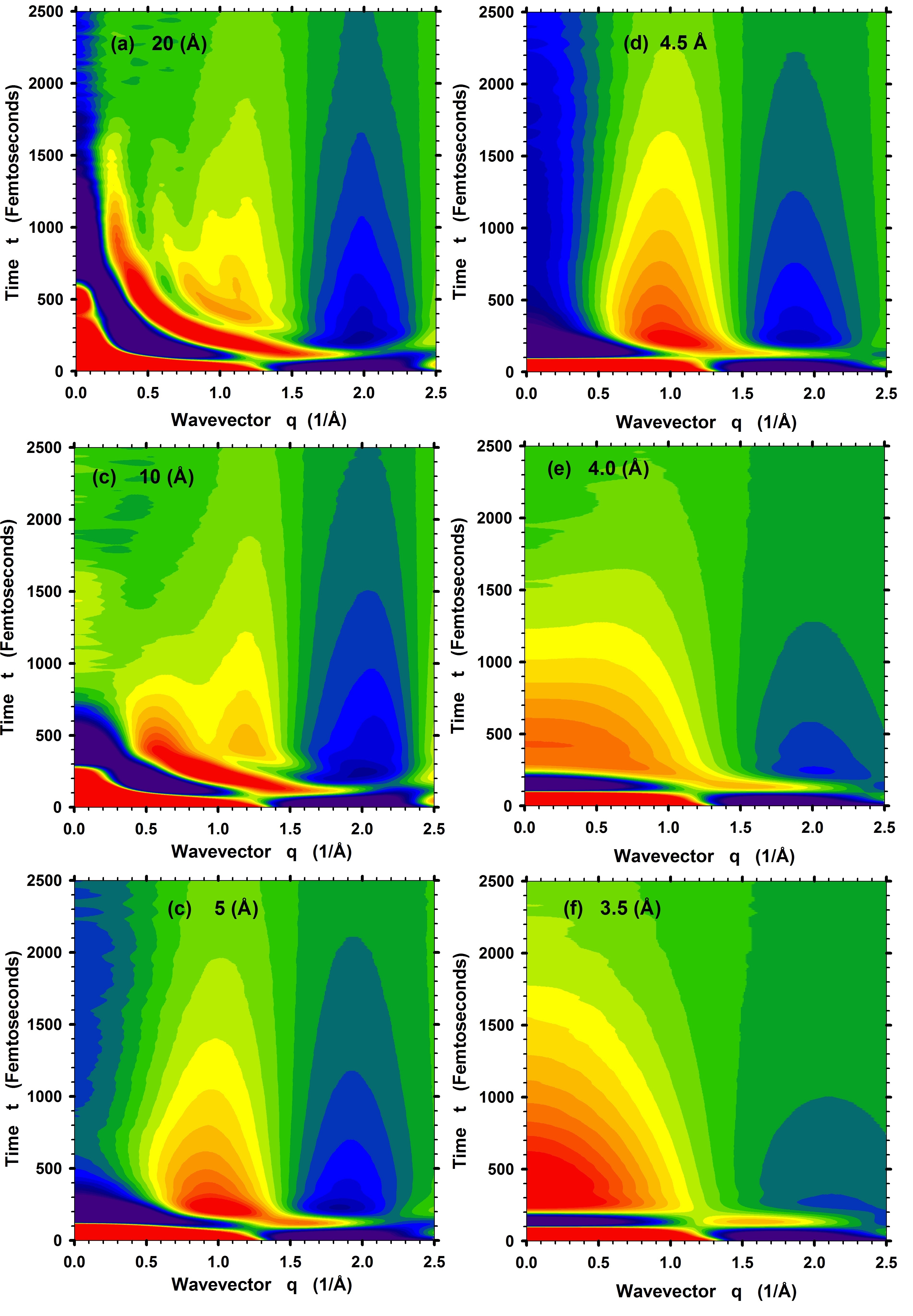}
\caption{Explanation of the \bonfire feature in the $r \rightarrow q$
Fourier transform of the \sscf. See panels (3,1) and (3,2) of Fig.\ref{fig:Fig1}. 
Different values of the upper cutoff, $R_{max}$,  
in the Fourier transform are shown in the panels.
}\label{fig:Fig11}
\end{center}
\end{figure}

\begin{figure}
\includegraphics[angle=0,width=3.4in]{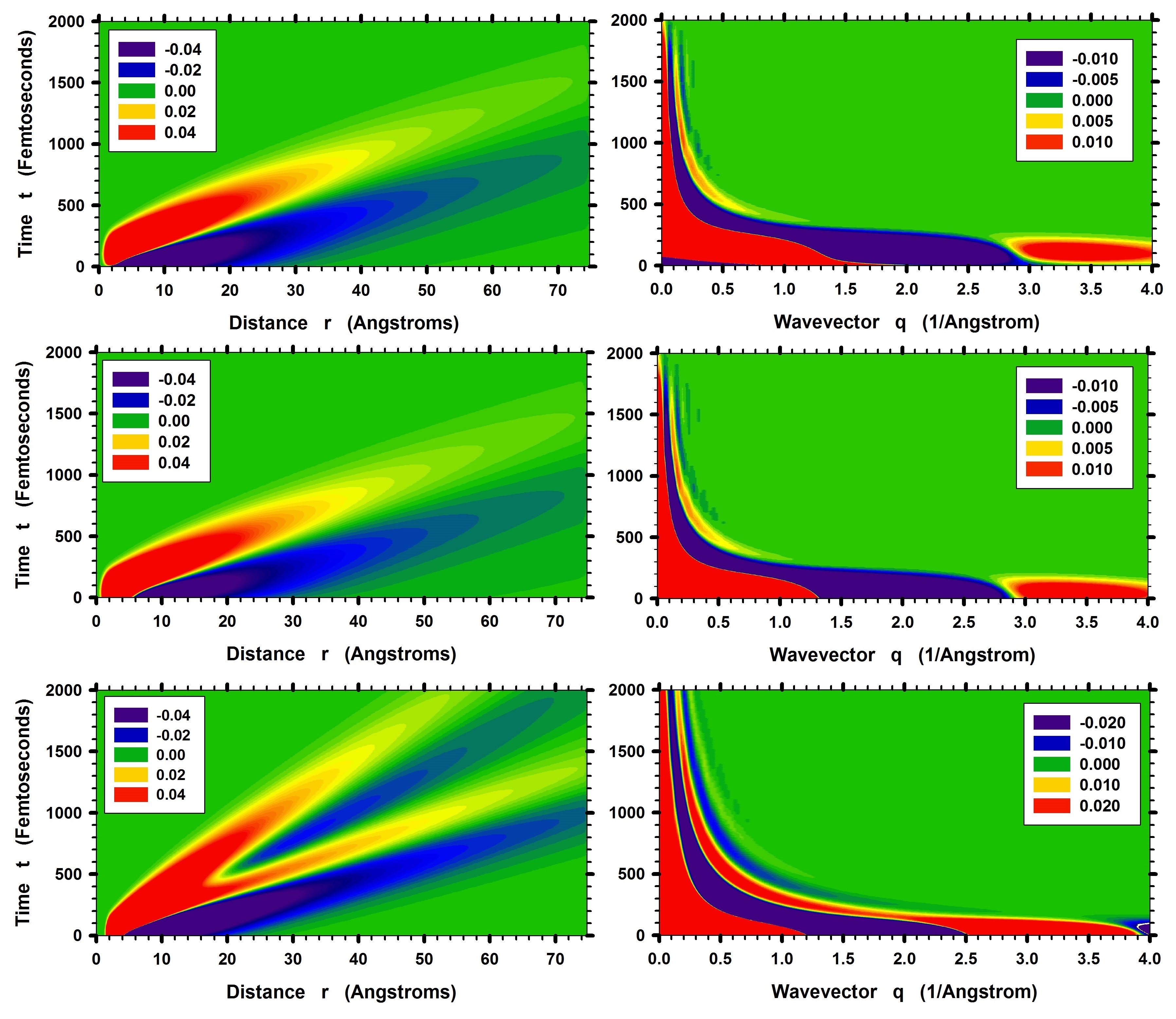}
\caption{Model \sscfs (on the left) and their Fourier transforms (on the right).
Upper 4 panels show that the \tongue feature is affected by the
wave's structure close to their origin.
Lower 4 panel demonstrates the overlap from the two waves in the Fourier transform.
See text for details. 
}\label{fig:Fig12}
\end{figure}

In order to understand the \bonfire feature in 
panels (3,1) and (3,2) of Fig.\ref{fig:Fig1}
we adopt an {\it ad hoc} approach. 
In particular, in performing the Fourier transforms, we
integrate from $r=0$ (\AA) up to some maximum value $R_{max}$ 
and check how the value of  $R_{max}$  affects the Fourier image. 
The results are shown in Fig.\ref{fig:Fig11}. 

Panel (a) shows the results of the integration up to $r=20$ (\AA). 
Note the similarities and differences between the intensity in panel (a) and 
the intensity in panel (3,2) of Fig.\ref{fig:Fig1}.
For panel (b) $R_{max}=10$ (\AA). 
By comparing panels (a) and (b) note
that the negative intensity region close to $q=0$ (\AA\textsuperscript{-1}) present in 
panel (a) is gone in panel (b).
Similarly gone is the part of the \sickle feature that apparently  
originates from the part of the shear stress wave that we do not count
when we integrate up to  $R_{max}=10$ (\AA).
However, the \bonfire feature is still there, even though
it is affected in the transition from (b) to (c).
In panel (c) the negative intensity region around  $q=0$ (\AA\textsuperscript{-1})
appears again. Thus, from the comparison of panels
(a), (b) and (c) we conclude that the negative intensity close
to $q=0$ (\AA\textsuperscript{-1} is related to the spatial extent of the \sscf.
Further note that the \bonfire feature is still present in panels (c) and (d).
The transition from panel (d) to panel (e) affects the \bonfire feature
very significantly. 
It follows from panel (a) of Fig.7 in Ref.\cite{Levashov2013} that 
the region between 4 (\AA) and 5 (\AA) corresponds to the interval of distances in which
the splitting of the second peak in the pair density function occurs. 
Overall, we conclude that the \bonfire feature is related to the absence of periodicity
in the \sscf for $r<10$  (\AA\textsuperscript{-1}). 
Comparison of panels (1,2) and (3,2) of Fig.\ref{fig:Fig1} also 
hints that this conclusion is correct. 
Thus, one may notice that the temporal extent of the
\bonfire feature in (3,2) corresponds to the temporal extent of
the bright vertical stripes in the region $r<10$ (\AA). 
The last idea can also be tested on the \sscfs and their
Fourier transforms at higher temperatures, as can be seen in
Fig.\ref{fig:Fig08},\ref{fig:Fig10}.
Thus a comparison of panels (1,1) and (1,3) of Fig.\ref{fig:Fig08}
shows that the bright stripe present in (1,1) at 5 (\AA) is significantly less pronounced 
in panel (1,3). 
The comparison of the corresponding panels in Fig.\ref{fig:Fig10}
shows that the \bonfire feature is present in (1,1), but nearly absent in (1,3).

\section{The \tongue feature \label{s:tongue}}

In order to understand the \tongue feature present in 
panels (3,1) and (3,2) of Fig.\ref{fig:Fig1}
we again adopt an {\it ad hoc} approach. 
Thus we create several model \sscfs and,
by comparing the Fourier transforms of these \sscfs, 
we demonstrate that the line shape of the \tongue feature 
is affected by the behavior of stress waves at distances 
at which the stress waves appear.

It is shown in Ref.\cite{Landau6} that in viscous liquids in spherically symmetric 
homogeneous cases the pressure profile far away from the origin
is given approximately by:
\begin{eqnarray}
p'(r,\tau)=p'_{o}\frac{\tau}{r^{5/2}}\exp\left[-\frac{\tau^2}{4ar}\right]\;\;,
\label{eq;ldaup-1}
\end{eqnarray}
where $p'(r,\tau)$ is the deviation of pressure from 
its average value in the system caused by the wave,
$\tau \equiv t-r/c$, $c$ - is the speed of the wave, 
$r$ is the distance from the origin of the wave, 
and $a$ controls the rate of the dissipation of the wave.
We use the functional form (\ref{eq;ldaup-1}) to create 
several model \sscfs. This does not mean that we assume that
(\ref{eq;ldaup-1})  correctly describes the shape of the shear stress waves.
However, we believe that this approach allows determination
of the origin of the \tongue feature.  

Panel (1,1) of Fig.\ref{fig:Fig12} shows the pressure profile 
calculated using (\ref{eq;ldaup-1}) with the following values of the parameters:
$c = 30$ (\AA/fs), $4a = 2500$ (fs\textsuperscript{2}/\AA), and $p'_{o} = 1$.
We use (\ref{eq;ldaup-1}) to calculate the pressure profile for $r>2$ (\AA).
We assume that for smaller distances the pressure is zero.
For smoothness we also convolute the function with the Gaussian function 
of width $\sigma_r = 0.50$ (\AA)  along the $r$-axis and with 
width $\sigma_t = 50$ (fs) along the $t$-axis. 

Panel (1,2) shows the Fourier transform of panel (1,1) into $(t,q)$-space.
We see in (1,2) the analogue of the \tongue feature and also the \sickle feature.
Note, however, that close to the origin the \tongue feature 
has a negative intensity. We found that this negative intensity could be removed by assuming that
the wave starts not at $r=0$ (\AA), but instead at some finite distance.
Thus, panel (2,1) of Fig.\ref{fig:Fig12} shows a pressure wave in which instead of
$\tau \equiv t-r/c$ we used $\tau = t - (r-r_b)/c$ with $r_b=3.3$ (\AA).
As before we assumed that for $r<2$ (\AA) the pressure is zero
and we convoluted the function with the parameters given above.
Panel (2,2) shows the Fourier transform of (2,1) in $(t,q)$-space.
Note that compared to panel (1,2) there is no negative intensity around the origin
in  panel (2,2).
The comparison also could be made with panel (2,1) 
of Ref.\cite{Levashov20141}.

Panel (3,1) shows a pressure profile which is a sum of two waves.
One wave is exactly the same as in panel (2,1). The second wave
also starts at $3.3$ (\AA) and has the following values of the parameters:
$c = 60$ (\AA/fs), $4a = 1000$ (fs\textsuperscript{2}/\AA) and $p'_{o} = 3$.
We see in panel (3,2) that the second wave makes the \sickle feature
much more pronounced. Note also the increased intensity and the increased width of the
positive intensity region around $q=0$ (\AA\textsuperscript{-1}). 

Thus we conclude that the line shape of the \tongue feature results from the behavior of 
the stress waves near their origin.



\begin{thebibliography}{10}
\bibitem{qmin}
The lowest value of $q$ follows from the condition $\exp(i q_{min}L) = 1$. 
Thus $q_{min}= 2\pi/L$. For the system with $(L/2) \approx 41.21$ (\AA) we have
$q_{min} \sim 0.076$ (\AA\textsuperscript{-1}). 
For the system with $(L/2) \approx 20.60$ (\AA) we have
$q_{min}=0.153$ (\AA\textsuperscript{-1}). 
Similarly, for the frequencies $\nu$. 
If our data span time interval from $t=0$ to $t_{max}$ then we should not consider frequencies smaller than
$\nu_{min}=1/t_{max}$. For $t_{max}=4000$ (fs) we have $\nu_{min} = 0.00025$ (fs\textsuperscript{-1}).
\end{thebibliography}
\end{document}